\documentclass[prd, twocolumn, superscriptaddress]{revtex4-2}
\usepackage{amsmath}
\usepackage{amssymb}

\usepackage{bm}

\usepackage{tikz}

\usepackage{array}

\allowdisplaybreaks[1]

\begin{document}
\title{Non-relativistic limit of scalar and Dirac fields in curved spacetime}

\author{Riccardo Falcone}
\affiliation{Department of Physics, University of Sapienza, Piazzale Aldo Moro 5, 00185 Rome, Italy}

\author{Claudio Conti}
\affiliation{Department of Physics, University of Sapienza, Piazzale Aldo Moro 5, 00185 Rome, Italy}
\affiliation{Institute for Complex Systems (ISC-CNR), Department of Physics, University Sapienza, Piazzale Aldo Moro 2, 00185, Rome, Italy}
\affiliation{Research Center Enrico Fermi, Via Panisperna 89a, 00184 Rome, Italy}

\begin{abstract}
We give from first principles the non-relativistic limit of scalar and Dirac fields in curved spacetime. We aim to find general relativistic corrections to the quantum theory of particles affected by Newtonian gravity, a regime nowadays experimentally accessible. We believe that the ever-improving measurement accuracy and the theoretical interest in finding general relativistic effects in quantum systems require the introduction of corrections to the Schr\"{o}dinger-Newtonian theory. We rigorously determine these corrections by the non-relativistic limit of fully relativistic quantum theories in curved spacetime. For curved static spacetimes, we show how a non-inertial observer (equivalently, an observer in the presence of a gravitational field) can distinguish a scalar field from a Dirac field by particle-gravity interaction. We study the Rindler spacetime and discuss the difference between the resulting non-relativistic Hamiltonians. We find that for sufficiently large acceleration, the gravity-spin coupling dominates over the corrections for scalar fields, promoting Dirac particles as the best candidates for observing non-Newtonian gravity in quantum particle phenomenology.
\end{abstract}

\maketitle

\section{Introduction}

The study of gravitational effects in quantum mechanics is driven by the search for a bridge between general relativity and the quantum theory. In the last twenty years, a remarkable series of experiments reported evidence of gravitational effects on the discrete spectrum of neutron bouncing~\cite{article1, PhysRevD.67.102002, article2, article3, article4, PhysRevLett.112.071101, Kamiya:2014qia}. These experiments confirmed the prediction of neutron wave functions having the form of Airy functions in the presence of an homogeneous gravity field.

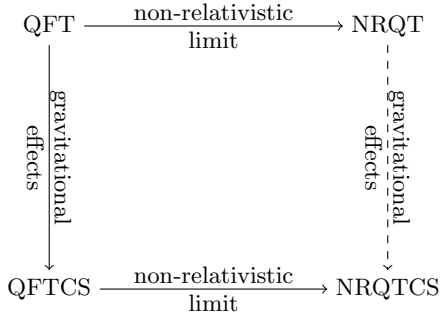
\begin{figure}
\begin{tikzpicture}
\node (QFT) [] {QFT};
\node (NRQT) [right of = QFT, node distance=4.5cm] {NRQT};
\node (QFTCS) [below of = QFT, node distance=3.5cm] {QFTCS};
\node (NRQTCS) [right of = QFTCS, node distance=4.5cm] {NRQTCS};

\path[->] (QFT) edge node[text width=2.5cm, anchor=center,align=center] {non-relativistic limit} (NRQT);
\path[->] (QFT) edge node[sloped, anchor=center, text width=2.5cm,align=center] {gravitational effects} (QFTCS);
\path[->] (QFTCS) edge node[text width=2.5cm, anchor=center,align=center] {non-relativistic limit} (NRQTCS);
\path[dashed,->] (NRQT) edge node[sloped, anchor=center, text width=2.5cm,align=center] {gravitational effects} (NRQTCS);
\end{tikzpicture}
\caption{The links between Quantum Field Theory (QFT), Non-Relativistic Quantum Theory (NRQT), Quantum Field Theory in a Curved Space Time (QFTCS), and its non-relativistic limit (NRQTCS). The path NRQT $\rightarrow$ NRQTCS is not rigorous as it ignores the relativistic nature of the fields.}\label{theory_limits}
\end{figure}

The reported observations can be explained by the Non-Relativistic Quantum Theory (NRQT) with an external gravitational Newtonian potential. This theoretical approach is the first step to analyze phenomena in the regime of Non-Relativistic Quantum Theory in Curved Spacetime (NRQTCS), ignoring the back-reaction of quantum particles on the gravitational field and any eventual quantum nature of gravity. In Fig.~\ref{theory_limits}, we represent the approach by two vertexes (NRQT and NRQTCS).

Despite being the most direct attack on the problem, the former approach can be inconsistent or too simplified. Indeed, the NRQT description of quantum particles approximates the fully-relativistic Quantum Field Theory (QFT) into a non-covariant theory. Therefore, in NRQT, we ignore the relativistic nature of fields. As a result, we may miss some interactions between matter and gravity arising from covariance, e.g., spin-gravity couplings for Dirac fields. A non-relativistic theory cannot furnish General Relativistic (GR) corrections. On the other hand, the experimental precision may eventually increase to the point that these GR corrections become detectable.

By looking at Fig.~\ref{theory_limits}, we identify these steps with the path QFT $\rightarrow$ NRQT $\rightarrow$ NRQTCS. The fully-relativistic QFT is approximated by NRQT in the non-relativistic limit, and, then, by considering gravitational effects, one studies the NRQTCS regime. The non-relativistic limit (QFT $\rightarrow$ NRQT) cancels out information \textit{before} the gravitational effects are introduced (NRQT $\rightarrow$ NRQTCS).

Another way to address the problem exists. Instead of introducing the gravitational effects \textit{after} the non-relativistic limit, we may consider them \textit{before} such a limit. In this way, we are able to take track of the GR corrections on the gravity-matter interaction avoiding the inconsistencies. The procedure relies on the Quantum Field Theory in Curved Spacetime (QFTCS), which is the description of fully-relativistic quantum fields affected by a gravitational field. QFTCS also ignores the back-reaction of the field on the metric (i.e., the gravity is not quantum), but it is the simplest attempt to a quantum theory that takes into account a non-flat metric. We identify the new approach in Fig. \ref{theory_limits} through the path QFT $\rightarrow$ QFTCS $\rightarrow$ NRQTCS, and corresponds to the non-relativistic limit of a fully-relativistic quantum field theory in a curved spacetime.

The most known predictions of QFTCS are the Hawking \cite{Hawking:1975vcx}, and Unruh effect \cite{PhysRevD.14.870}, which have never been directly observed due to their inaccessible energy scales. Conversely, the neutron-bouncing experiments \cite{article1, PhysRevD.67.102002, article2, article3, article4, PhysRevLett.112.071101, Kamiya:2014qia} prove that the NRQTCS regime is nowadays experimentally accessible. This circumstance motivates the study of the non-relativistic limit of QFTCS. For instance, in a recent work \cite{2021IJMPD..3050098R}, the problem of quantum bouncing particles in a gravitational field is discussed in the context of QFTCS. By solving the Dirac equation in Rindler spacetime with bouncing boundary conditions, the authors found GR corrections to the energy spectrum of the neutrons in a gravitational field. Others considered related scenarios: authors in \cite{PhysRevLett.44.1559, PhysRevD.22.1922, PhysRevD.25.3180} found the perturbations of the energy levels of an atom placed in curved spacetime; in \cite{PhysRevA.88.022121}, a generalized Schwarzschild metric is used to investigate GR corrections with gravitational spin-orbit coupling. These results were derived from solving the Dirac equation in Rindler spacetime in a non-relativistic limit and, hence, by following the path QFTCS $\rightarrow$ NRQTCS.

Here, we report on a general procedure to perform the non-relativistic limit for bosonic and fermionic fields in a static spacetime. We consider complex scalar and Dirac fields and provide the non-relativistic description of quantum particles in terms of wave functions, scalar product, and Hamiltonian.

It is known that, in the Minkowski spacetime, the time evolution of free non-relativistic single-particles can be approximately described by the free Hamiltonian, which has the same form for both scalar and Dirac fields. Indeed, the Klein-Gordon and the Dirac equation asymptotically lead to the same non-relativistic Schr\"{o}dinger equation. For a Dirac field, the spinorial components --- obeying the same Schr\"{o}dinger equation --- are decoupled and can be treated as spectral degeneracy. Therefore, without a spin-dependent interaction or enough experimental precision, a Minkowski observer cannot distinguish the time evolution of a non-relativistic scalar particle from a Dirac particle. This also happens if one introduces a first-order correction due to a weak gravitational field. In the case of a Rindler spacetime with a nearly flat metric and for non-relativistic particles, the first correction introduced in the Schr\"{o}dinger equation corresponds to the Newtonian gravitational potential, with no difference between scalar and Dirac field.

By considering GR corrections, the difference between scalar and Dirac fields appears. In this manuscript, we show that metrics not approximated by the flat spacetime lead to a non-vanishing difference between the Schr\"{o}dinger equations arising from the Klein-Gordon and Dirac equation in curved spacetime. A spin-metric coupling occurs, and the observer can distinguish between a scalar and a Dirac particle. We also show that for approximately flat metrics, such coupling can be observed at different orders. For sufficiently large curvature, the precision required to distinguish between scalar and Dirac fields is lower than the one needed in flat spacetime.

The paper is organized as follows. In Sec.~\ref{Minkowski_spacetime} we give a review for the non-relativistic limit of scalar [Sec.~\ref{Minkowski_spacetime_Scalar_field}] and Dirac [Sec.~\ref{Minkowski_spacetime_Dirac_field}] fields in the Minkowski spacetime. We also show how non-relativistic particles are approximately solutions to the same Schr\"{o}dinger equation. Sec.~\ref{NonMinkowksi_spacetime}, is devoted to the curved case. We derive the non-relativistic limit of fields in a static spacetime and we show how the approximated Schr\"{o}dinger equations differ in the two cases. Finally, we detail these results for the case of Rindler metric in Sec.~\ref{Rindler_frame}. Conclusions are drawn in Sec.~\ref{Conclusions}.

\section{Minkowski spacetime}\label{Minkowski_spacetime}

In the present section, we work in a Minkowski spacetime, defined by coordinates $(t,\vec{x})$ and flat metric
\begin{equation}
\eta_{\mu\nu} = \text{diag}(-c^2,1,1,1),
\end{equation}
where $c$ is the speed of light. We consider a complex scalar $\hat{\phi}(t,\vec{x})$ and Dirac $\hat{\psi}(t,\vec{x})$ field. We review the known description of fields in terms of particles --- see for instance \cite{Wald:1995yp} --- and we detail the non-relativistic limit, identified by states with kinetic and potential energy small with respect to their mass energy.

We show that the representation of particles through positive-frequency solutions of the Klein-Gordon (Dirac) equation and the associated scalar product leads to the familiar position representation of $0$-($1/2$-)spin states in the non-relativistic limit. We also show that the time-evolution of these states leads to the Schr\"{o}dinger equation in NRQT. In absence of spin-dependent interaction, scalar and Dirac particles are approximately described by the same Hamiltonian and, hence, identical
in their time evolution.

\subsection{Scalar field}\label{Minkowski_spacetime_Scalar_field}

In the case of a scalar field, we start reviewing free particles --- i.e. without interaction --- and we use the decomposition in positive and negative frequency modes with fixed momenta. Positive frequency modes are a basis for the Hilbert space of single-particles, and the Klein-Gordon product is adopted as the inner product. By considering the non-relativistic limit, we show that these modes lead to the position representation of particles with fixed momenta and the Klein-Gordon scalar product can be approximated by the usual $L^2(\mathbb{R}^3)$ inner product of NRQT. Moreover, we show that the evolution of these states can be approximated by the free Schr\"{o}dinger equation. In this way, we recover the non-relativistic description of free particles in terms of wave functions, scalar product and free Hamiltonian.

Then, we derive the same description of non-relativistic particles starting from a general decomposition of the field in terms of positive and negative frequency modes. These new modes are not necessarily associated to particles with fixed momenta and they lead to the position representation of states with fixed quantum numbers. Also, we estimate the errors for the non-relativistic approximations.

Finally, we describe the interacting case by a non-vanishing external potential. We adopt the interaction picture and define particles states as time-dependent combinations of free particles. In the Schr\"{o}dinger picture, we represent a generic particle state as a time-dependent combination of free-evolving modes. We show that in the non-relativistic limit, these wave functions are approximated solutions of a Schr\"{o}dinger equation with a potential. Also, we show that the product of two single-particle states can be approximated by the $L^2(\mathbb{R}^3)$ product of their wave functions.

As anticipated, we start from considering a free complex scalar field $\hat{\phi}$ solution of the Klein-Gordon equation
\begin{equation} \label{Klein_Gordon}
\left[ c^2 \eta^{\mu\nu} \partial_\mu \partial_\nu - \left( \frac{mc^2}{\hbar} \right)^2 \right] \hat{\phi} = 0,
\end{equation}
where $\eta^{\mu\nu}$ is the inverse of $\eta_{\mu\nu}$ and $m$ the mass. Hereafter, we adopt Einstein notation over repeated indexes. Greek indexes $\mu, \nu, \rho, \sigma$ are for $4$ dimensional spacetime coordinates $(0,1,2,3) = (t, \vec{x})$, while Latin indexes $i, j, k$ for $3$ dimensional space coordinates $(1,2,3) = \vec{x}$.

Equation (\ref{Klein_Gordon}) leads to the usual expression for the scalar field
\begin{equation} \label{free_field}
\hat{\phi}(t,\vec{x}) = \int d^3 k \left[ f(\vec{k},t,\vec{x}) \hat{a}(\vec{k}) + f^*(\vec{k},t,\vec{x}) \hat{b}^\dagger(\vec{k}) \right],
\end{equation}
where $\hat{a}(\vec{k})$ is the annihilator operator for a free particle mode
\begin{equation}\label{free_modes}
f(\vec{k},t,\vec{x}) =  \sqrt{\frac{\hbar c^2}{(2\pi)^3 2 \omega(k)}} e^{-i\omega(k)t + i\vec{k} \cdot \vec{x}},
\end{equation}
$\hat{b}^\dagger(\vec{k})$ is the creation operator for an antiparticle with momentum $\vec{k}$ and $\omega(k)$ is the dispersion relation
\begin{equation} \label{dispersion_relation}
\omega(k) = \sqrt{\left(\frac{mc^2}{\hbar}\right)^2 + (c k)^2}.
\end{equation}
$\hat{a}(\vec{k})$ and $\hat{b}(\vec{k})$ generate the usual Minkowski-Fock space through the canonical commutation relation
\begin{subequations}
\begin{align}
& [\hat{a}(\vec{k}),\hat{a}(\vec{k'})] = [\hat{b}(\vec{k}),\hat{b}(\vec{k'})] = [\hat{a}(\vec{k}),\hat{b}(\vec{k'})] = 0,\\
& [\hat{a}(\vec{k}),\hat{a}^\dagger(\vec{k'})] = [\hat{b}(\vec{k}),\hat{b}^\dagger(\vec{k'})] = \delta^3(\vec{k}-\vec{k}').
\end{align}
\end{subequations}

The $f(\vec{k})$ modes are defined to be solutions of the Klein-Gordon equation (\ref{Klein_Gordon}) and orthonormal with respect to the Klein-Gordon scalar product
\begin{align}\label{KG_scalar_product}
( \phi, \phi' )_\text{KG} = & \frac{i}{\hbar c^2} \int_{\mathbb{R}^3} d^3x \left[ \phi^*(t,\vec{x}) \partial_0  \phi'(t,\vec{x}) \right. \nonumber \\
& \left. -  \phi'(t,\vec{x}) \partial_0 \phi^*(t,\vec{x}) \right],
\end{align}
which is defined for any $t$ and for any $\phi $, $ \phi'$ solutions of Eq.~(\ref{Klein_Gordon}). Equation (\ref{KG_scalar_product}) is time-independent for such solutions, as it can be directly proven by using Eq.~(\ref{Klein_Gordon}) and the integration by parts:
\begin{align}
\frac{d}{dt} ( \phi, \phi' )_\text{KG} = & \frac{i}{\hbar c^2} \int_{\mathbb{R}^3} d^3x \left( \phi^* \partial^2_0  \phi'  -  \phi' \partial^2_0 \phi^* \right)\nonumber \\
 = & \frac{i}{\hbar c^2} \int_{\mathbb{R}^3} d^3x \left( \phi^* \delta^{ij} \partial_i \partial_j \phi' -  \phi' \delta^{ij} \partial_i \partial_j \phi^* \right)\nonumber \\
 = & \frac{i \delta^{ij} }{\hbar c^2} \int_{\mathbb{R}^3} d^3x \left[ -(\partial_i \phi^*) (\partial_j \phi') +  (\partial_i \phi') (\partial_j \phi^*) \right]\nonumber \\
 = & 0,
\end{align}
where $\delta^{ij}=\eta^{ij}$ is the Kronecker delta. The orthonormality of $f(\vec{k})$ modes with respect to $( \phi, \phi' )_\text{KG}$ reads
\begin{subequations}\label{KG_scalar_product_orthonormality_f}
\begin{align}
& ( f(\vec{k}), f(\vec{k}') )_\text{KG} = \delta^3(\vec{k}-\vec{k}'), \\
& ( f^*(\vec{k}), f^*(\vec{k}') )_\text{KG} = -\delta^3(\vec{k}-\vec{k}'),\\
& ( f(\vec{k}), f^*(\vec{k}') )_\text{KG} = 0,
\end{align}
\end{subequations}
which can be proven from Eqs.~(\ref{free_modes}) and (\ref{KG_scalar_product}).

In the interaction-free theory, the Hilbert space of single-particles is the vector space generated by the $f(\vec{k})$ modes and supplemented by the Klein-Gordon scalar product (\ref{KG_scalar_product}). The $f^*(\vec{k})$ modes have to be excluded, since they are associated to negative probabilities [Eq.~(\ref{KG_scalar_product_orthonormality_f})]. The space of single antiparticle states is analogously defined from the field $\hat{\phi}^\dagger$. Once the single-particle and antiparticle space is defined, one can derive their Fock space, which is regarded as the space of the field states. Hereafter, we only focus on particles.

Each $f(\vec{k})$ mode is associated to a single-particle state $f(\vec{k}) \mapsto | \vec{k} \rangle = \hat{a}^\dagger(\vec{k}) | 0_\text{M} \rangle$ --- with $| 0_\text{M} \rangle$ as the vacuum state --- while the function $f(\vec{k},t,\vec{x})$, with varying $t$ and $\vec{x}$, provides a representation for $| \vec{k} (t) \rangle$, evolved with respect to the free theory
\begin{equation}\label{free_evolution_k_states}
i \hbar \partial_0 | \vec{k} (t) \rangle = \omega(\vec{k}) | \vec{k} (t) \rangle.
\end{equation}
We will show that in the non-relativistic limit, such representation can be approximated by the familiar position representation in NRQT.

The $| \vec{k} \rangle$ states are a basis for the single-particle space. This means that a generic particles state $| \phi \rangle$ can be expanded in the following way
\begin{equation}\label{free_state_decomposition}
| \phi \rangle  = \sum_{n=0}^\infty \int_{\mathbb{R}^{3n}} d^{3n} \textbf{k}_n \tilde{\phi}_n (\textbf{k}_n) | \textbf{k}_n \rangle,
\end{equation}
where
\begin{equation}
\textbf{k}_n = (\vec{k}_1, \dots, \vec{k}_n)
\end{equation}
is a $3n$ vector collecting $n$ momenta, $| \textbf{k}_n \rangle$ the $n$-particles state with momenta $\vec{k}_1, \dots, \vec{k}_n$, $| \textbf{k}_0 \rangle = | 0_\text{M} \rangle$ the vacuum state, $\tilde{\phi}_n (\textbf{k}_n)$ the $n$-particles wave function of $| \phi \rangle$ in the momentum representation, $\tilde{\phi}_0$ the coefficient associated to the vacuum state. $\tilde{\phi}_n (\textbf{k}_n)$ is defined to be symmetric with respect to the momenta variables. The representative of the time-evolved state $| \phi (t) \rangle$ in the Schr\"{o}dinger picture reads
\begin{align} \label{free_wave_function}
\phi_n (t, \textbf{x}_n) = & \left( \frac{2 m}{\hbar^2} \right)^{n/2} \int_{\mathbb{R}^{3n}} d^{3n} \textbf{k}_n \tilde{\phi}_n (\textbf{k}_n) \prod_{l=1}^n f(\vec{k}_l, t, \vec{x}_l),
\end{align}
where
\begin{equation}
\textbf{x}_n = (\vec{x}_1, \dots, \vec{x}_n).
\end{equation}
When $n=0$, we assume $\phi_0 = \tilde{\phi}_0$.

The Klein-Gordon scalar product (\ref{KG_scalar_product}) represents the Hilbert product between two single-particle states in terms of their wave functions
\begin{equation} \label{scalar_product_representation}
\langle \phi | \phi' \rangle = \frac{\hbar^2}{2m} ( \phi_1, \phi'_1 )_\text{KG}.
\end{equation}
This product is time independent if evaluated on wave functions of the form of Eq.~(\ref{free_state_decomposition}) and, hence, leading to constant probabilities, as expected by the quantum theory. Equation (\ref{scalar_product_representation}) can be proven by using Eqs.~(\ref{KG_scalar_product_orthonormality_f}), (\ref{free_state_decomposition}), (\ref{free_wave_function}) and the orthonormality of momentum states $\langle \vec{k} | \vec{k}' \rangle = \delta(k-k')$.

The time evolution of the $| \vec{k} \rangle$ states is described by the $e^{-i \omega t}$ phase of the time evolved $f(\vec{k},t,\vec{x})$ modes [Eq.~(\ref{free_evolution_k_states})]. In other words, the $| \vec{k} \rangle$ states are eigenstates of an Hamiltonian $\hat{h}_\text{KG}$ with $\omega(\vec{k})$ as eigenvalue. If we try to represents such Hamiltonian in the representation space of $f(\vec{k})$ modes, we must rely on some kind of square root of
\begin{equation}\label{H_KG}
H_\text{KG} = - (\hbar c)^2 \delta^{ij}\partial_i \partial_j + (mc^2)^2,
\end{equation}
since each $f(\vec{k})$ mode is solution of 
\begin{equation}
H_\text{KG} f(\vec{k}) = [\hbar \omega(\vec{k})]^2 f(\vec{k}).
\end{equation}

What we mean by square root of $H_\text{KG}$ is the fact that $H_\text{KG}$ and the representative of $\hat{h}_\text{KG}$ share the same eigenvectors in the $f(\vec{k})$ modes space, but with different eigenvalues: if $\hbar \omega$ is the eigenvalue of $| \vec{k} \rangle$ with respect to $\hat{h}_\text{KG}$, then $(\hbar \omega)^2$ is the eigenvalue of $f(\vec{k})$ with respect to $H_\text{KG}$. We define, in any case, the representative of $\hat{h}_\text{KG}$ as $h_\text{KG}$ and we write the following improper expression
\begin{equation}\label{h_KG_H_KG}
h_\text{KG} = \sqrt{H_\text{KG}}.
\end{equation}

$h_\text{KG}$ is not in a standard form, as it cannot be written in terms of spatial derivatives and space-dependent functions. However, in the non-relativistic limit, we show that $h_\text{KG}$  is approximated by the usual free-particle Hamiltonian that includes a mass and kinetic energy term
\begin{equation}
H_\text{M} = m c^2  - \frac{\hbar^2}{2 m} \delta^{ij}\partial_i \partial_j.
\end{equation}
This can be done by showing that the free modes $f(\vec{k})$ are approximately solutions of the Schr\"{o}dinger equation with $H_\text{M} $ as Hamiltonian
\begin{equation} \label{Schrodinger_free_mass}
i \hbar \partial_0 f(\vec{k}) \approx H_\text{M} f(\vec{k}).
\end{equation}

To do so, we remark that the non-relativistic limit is achieved by particle states with energies very close to the mass energy
\begin{equation}\label{non_relativistic_limit}
\left| \frac{\hbar \omega}{mc^2} - 1 \right| \ll 1.
\end{equation}
We say that $| \phi \rangle $ is non-relativistic if $\tilde{\phi}_n (\textbf{k}_n)$ is non-vanishing only for momenta $\vec{k}$ such that Eq.~(\ref{non_relativistic_limit}) holds.

For non-relativistic momenta $\vec{k}$, the frequency dispersion relation of Eq.~(\ref{dispersion_relation}) can be approximated by
\begin{equation} \label{dispersion_relation_approximation}
\omega(k) \approx \frac{m c^2}{\hbar} + \frac{\hbar k^2}{2m},
\end{equation}
and, hence, $f(\vec{k},t,\vec{x})$ reads
\begin{equation}\label{Minkowski_modes_nonrelativistic}
f(\vec{k},t,\vec{x}) \approx \frac{\hbar}{\sqrt{(2\pi)^3 2 m}} \exp \left( -i\frac{m c^2 t}{\hbar} - i \frac{\hbar k^2 t}{2m} + i\vec{k} \cdot \vec{x} \right).
\end{equation}
This means that $f(\vec{k})$ is approximately solution to Eq.~(\ref{Schrodinger_free_mass}). Moreover, it is possible to notice that in Eq.~(\ref{Minkowski_modes_nonrelativistic}) the mode $(\sqrt{2m}/\hbar) f(\vec{k},t,\vec{x})$ is put in the form of the familiar wave function of a momentum state $|\vec{k} \rangle$ in the position representation.

The fact that non-relativistic modes $f(\vec{k})$ are solutions of Eq.~(\ref{Schrodinger_free_mass}) means also that any time-dependent wave function in the Schr\"{o}dinger picture [Eq.~(\ref{free_wave_function})] is solution of the Schr\"{o}dinger equation for Fock states
\begin{equation}\label{Schrodinger_free_mass_n_particles}
i \hbar \partial_0  \phi_n   \approx \sum_{l=1}^n \left( m c^2  - \frac{\hbar^2}{2 m} \nabla^2_{\vec{x}_l} \right) \phi_n,
\end{equation}
where
\begin{equation}
\nabla^2_{\vec{x}} = \delta^{ij}\frac{\partial}{\partial x^i}\frac{\partial}{\partial x^j} .
\end{equation}
In this way, we describe the time evolution of any non-relativistic state in the familiar Schr\"{o}dinger picture of NRQT.

It can be noticed that $H_\text{M}$ is hermitian with respect to both the Klein-Gordon scalar product (\ref{KG_scalar_product}) and the $L^2(\mathbb{R}^3)$ inner product, defined as
\begin{equation}\label{L2_inner_product}
( \phi, \phi' )_{L^2(\mathbb{R}^3)} = \int_{\mathbb{R}^3} d^3x \phi^*(t,\vec{x}) \phi'(t,\vec{x}).
\end{equation}
Indeed, by integrating by parts, one can prove that
\begin{subequations}
\begin{align}
& ( H_\text{M} \phi,  \phi' )_\text{KG}  = ( \phi, H_\text{M}  \phi' )_\text{KG}, \\
 & ( H_\text{M} \phi,  \phi' )_{L^2(\mathbb{R}^3)}  = ( \phi, H_\text{M}  \phi' )_{L^2(\mathbb{R}^3)}.
\end{align}
\end{subequations}

It is straightforward to prove from Eq.~(\ref{Minkowski_modes_nonrelativistic}) that for non-relativistic modes, the Klein-Gordon scalar product (\ref{KG_scalar_product}) can be approximated by the $L^2(\mathbb{R}^3)$ inner product, with the exception of a $2 m / \hbar^2$ factor
\begin{equation}\label{KG_scalar_product_f_nonrelativistic}
( f(\vec{k}), f(\vec{k}') )_\text{KG} \approx \frac{2m}{\hbar^2} ( f(\vec{k}), f(\vec{k}') )_{L^2(\mathbb{R}^3)}. 
\end{equation}
By using Eqs.~(\ref{free_wave_function}) and (\ref{scalar_product_representation}), we can derive the same approximation for non-relativistic single particle states
\begin{equation}\label{KG_scalar_product_single_particle_nonrelativistic}
\langle \phi | \phi' \rangle \approx ( \phi_1 , \phi'_1 )_{L^2(\mathbb{R}^3)}. 
\end{equation}
This approximation can be also generalized for the case of an indefinite number of particles
\begin{equation}\label{KG_scalar_product_nonrelativistic}
\langle \phi | \phi' \rangle \approx \sum_{n=0}^\infty( \phi_n , \phi'_n )_{L^2(\mathbb{R}^{3n})},
\end{equation}
where
\begin{subequations}
\begin{align}
& ( \phi_n , \phi'_n )_{L^2(\mathbb{R}^{3n})} = \int_{\mathbb{R}^{3n}} d^{3n} \textbf{x}_n \phi^*_n (t, \textbf{x}_n)  \phi'_n (t, \textbf{x}_n), \\
& ( \phi_0 , \phi'_0 )_{L^2(\mathbb{R}^0)} = \phi_0^* \phi'_0.
\end{align}
\end{subequations}

\begin{table}
\begin{tabular}{c|c|c|}
 & QFT & NRQT \\
\hline
$\langle \phi | \phi' \rangle $ & $[\hbar^2/(2m)] ( \phi_1, \phi'_1 )_\text{KG}$ & $ ( \phi_1 , \phi'_1 )_{L^2(\mathbb{R}^3)}$\\
\hline
Hamiltonian & $h_\text{KG}$ & $H_\text{M}$ \\
\hline
\end{tabular}
\caption{Inner product (first line) and Hamiltonian (second line) for free scalar single-particles. The left column is for the fully relativistic theory (QFT), while the right one is for the non-relativistic limit (NRQT).} \label{table_scalar_Minkowski}
\end{table}

While in the fully relativistic theory, single-particles can be described by the inner product (\ref{scalar_product_representation}) and the Hamiltonian $h_\text{KG}$, non-relativistic single-particles can be approximately described by the $L^2(\mathbb{R}^3)$ inner product and the Hamiltonian $H_\text{M}$. This difference is shown schematically by Table \ref{table_scalar_Minkowski}. The Schr\"{o}dinger equation (\ref{Schrodinger_free_mass_n_particles}) and the inner product (\ref{KG_scalar_product_nonrelativistic}) are the familiar ingredients for the description of free Fock states in the position representation. In this way, we have been able to describe free scalar particles in NRQT, through the usual prescription.

It can be noticed that, in order to obtain Eqs.~(\ref{Schrodinger_free_mass}) and (\ref{KG_scalar_product_f_nonrelativistic}), we have used the explicit form of free modes $f(\vec{k})$ and performed the non-relativistic limit for such functions. Conversely, it is possible to show that the Klein-Gordon equation leads to a free Schr\"{o}dinger equation and the Klein-Gordon product (\ref{KG_scalar_product}) to an $L^2(\mathbb{R}^3)$ product for modes with positive frequencies without looking at the explicit form of such modes. The result is the same shown by Eqs.~(\ref{Schrodinger_free_mass_n_particles}) and (\ref{KG_scalar_product_nonrelativistic}). However, the method relies on a general definition of real frequency modes.

To see this, we expand the scalar field $\hat{\phi}$ in terms of generic modes $g(\theta)$ and $h(\theta)$ with, respectively, positive and negative frequencies:
\begin{equation}\label{free_field_positive_negative_frequencies}
\hat{\phi}(t,\vec{x}) = \sum_\theta \left[ g(\theta, t,\vec{x}) \hat{a}(\theta)  + h(\theta, t,\vec{x}) \hat{b}^\dagger(\theta) \right],
\end{equation}
where $\theta$ is a collection of quantum numbers which can be discrete, continuum or both. $ \sum_\theta$ is, hence, a generalized sum, including eventually integrals for continuum variables. $\hat{a}(\theta)$ and $\hat{b}^\dagger(\theta)$ are, respectively, annihilation operator for particle mode $g(\theta)$ and creation operator for antiparticle mode $h^*(\theta)$. The function $g(\theta, t,\vec{x})$  with varying $t$ and $\vec{x}$ is, hence, the representative of the single-particle state $|\theta \rangle $ with quantum numbers $\theta$. 

The fact that $g(\theta)$ and $h(\theta)$ have positive and negative frequencies can be expressed by the following time-dependencies:
\begin{subequations}\label{positive_negative_frequencies}
\begin{align}
& g(\theta,t,\vec{x}) = \tilde{g}_n(\theta,\vec{x}) e^{-i \omega(\theta) t}, \label{positive_frequencies}\\
& h(\theta,t,\vec{x}) = \tilde{h}_n(\theta,\vec{x}) e^{i \omega(\theta) t},
\end{align}
\end{subequations}
where the function $\omega(\theta)$ is many-to-one because of the energy degeneracy. The orthonormality with respect to the Klein-Gordon scalar product (\ref{KG_scalar_product}), instead, reads
\begin{subequations}\label{KG_scalar_product_orthonormality_g}
\begin{align}
& ( g(\theta), g(\theta') )_\text{KG} = \delta_{\theta\theta'}, \\
 & ( h(\theta), h(\theta') )_\text{KG} = -\delta_{\theta\theta'}, \\
  & ( g(\theta), h(\theta') )_\text{KG} = 0,
\end{align}
\end{subequations}
where, in this case, the deltas are generalized, as they act as Kronecker deltas for discrete indexes and as Dirac deltas for continuum variables.

The decomposition of the field in real frequencies [Eq.~(\ref{positive_negative_frequencies})] is guaranteed by the Klein-Gordon equation (\ref{Klein_Gordon}). Indeed, by imposing the ansatz (\ref{positive_frequencies}), Eq.~(\ref{Klein_Gordon}) for $g(\theta)$ becomes a Schr\"{o}dinger equation with eigenvalues proportional to $\omega^2$
\begin{align} \label{Klein_Gordon_Schrodinger}
H_\text{KG} g(\theta) = [\hbar \omega(\theta)]^2 g(\theta).
\end{align}
$H_\text{KG}$ is positive with respect to the Klein-Gordon scalar product (\ref{KG_scalar_product}) for any positive-frequency solution of the Klein-Gordon equation. Indeed, by defining
\begin{align}
& h_0 = mc^2 , & h_i =\hbar c \partial_i,
\end{align}
one can prove, through integration by parts, that
\begin{equation}
( \phi, H_\text{KG} \phi' )_\text{KG} = \delta^{i j} ( h_i \phi, h_j \phi' )_\text{KG} + ( h_0 \phi, h_0 \phi' )_\text{KG}.
\end{equation}
In this way one can see that if $\phi$ is combination of $g(\theta)$ modes, then
\begin{equation}
( \phi, H_\text{KG} \phi )_\text{KG} > 0
\end{equation}
and, hence, $H_\text{KG}$ has positive eigenvalues in the space of $g(\theta)$ modes. This is compatible with the fact that the $\omega$ appearing in Eq.~(\ref{Klein_Gordon_Schrodinger}) is real. The same proof holds for $h^*(\theta)$ modes, by considering the field $\hat{\phi}^\dagger$.

As in Eqs.~(\ref{free_state_decomposition}) and (\ref{free_wave_function}), we may define wave functions for any state $|\phi \rangle$ by decomposing it in terms of $| \theta \rangle$ states:
\begin{align} \label{general_Fock_expansion}
| \phi \rangle  = & \sum_{n=0}^\infty \sum_{\bm{\theta}_n} \tilde{\phi}_n (\bm{\theta}_n) | \bm{\theta}_n \rangle,
\end{align}
where we have defined the vector
\begin{equation}
\bm{\theta}_n = (\theta_1, \dots, \theta_n).
\end{equation}
$\tilde{\phi}_n (\bm{\theta}_n)$ is symmetric with respect to $\theta_1, \dots, \theta_n$. The state $| \phi \rangle$ in the Schr\"{o}dinger picture is represented by
\begin{equation}\label{wavefunction_g}
\phi_n (t, \textbf{x}_n) =  \left( \frac{2 m}{\hbar^2} \right)^{n/2}  \sum_{\bm{\theta}_n} \tilde{\phi}_n (\bm{\theta}_n)  \prod_{l=1}^n g(\theta_l,t,\vec{x}_l).
\end{equation}

We prove that in the non-relativistic limit (\ref{non_relativistic_limit}), $\phi_n$ is approximately solution of the free Schr\"{o}dinger equation (\ref{Schrodinger_free_mass_n_particles}) by showing that $g(\theta)$ is approximately solution of the free single-particle Schr\"{o}dinger equation (\ref{Schrodinger_free_mass}):
\begin{equation} \label{Schrodinger_free_mass_positive_negative_frequencies}
i \hbar \partial_0 g(\theta) \approx H_\text{M} g(\theta).
\end{equation}
Thanks to Eqs.~(\ref{wavefunction_g}) and (\ref{Schrodinger_free_mass_positive_negative_frequencies}), one can check that Eq.~(\ref{Schrodinger_free_mass_n_particles}) holds also for wave functions defined by Eq.~(\ref{wavefunction_g}). 

The proof of Eq.~(\ref{Schrodinger_free_mass_positive_negative_frequencies}) follows from the fact that $g(\theta)$ is solution of Eq.~(\ref{Klein_Gordon}) and in the non-relativistic limit (\ref{non_relativistic_limit}), the second-order time derivative of Eq.~(\ref{Klein_Gordon}) acting on $g(\theta)$ is approximately replaced by a first-order time derivative. Indeed, by using Eq.~(\ref{positive_frequencies}), we obtain the following chain of identities
\begin{align} \label{Klein_Gordon_approximation_positive_negative_frequencies}
- \partial_0^2 g(\theta) = & \omega^2(\theta)  g(\theta) \nonumber \\
 = & \left( \frac{mc^2}{\hbar} \right)^2 \left\lbrace 1 + \left[ \frac{\hbar \omega(\theta)}{mc^2} - 1 \right] \right\rbrace^2 g(\theta)  \nonumber \\
 = & \left( \frac{mc^2}{\hbar} \right)^2\left\lbrace 1 + 2 \left[ \frac{\hbar \omega(\theta)}{mc^2}  - 1 \right] + \mathcal{O}(\epsilon^2) \right\rbrace g(\theta) \nonumber \\
= &  \frac{mc^2}{\hbar}\left[ 2 i \partial_0 - \frac{mc^2}{\hbar} + \frac{mc^2}{\hbar} \mathcal{O}(\epsilon^2) \right] g(\theta),
\end{align}
with
\begin{equation}\label{non_relativistic_order}
\epsilon =\frac{\hbar \omega}{mc^2} - 1 .
\end{equation}
Hereafter we do not specify the argument of $\epsilon$ since for different non-relativistic frequencies $\omega$, $\omega'$, we have that $\epsilon(\omega) \sim \epsilon(\omega')$. Finally, by using Eq.~(\ref{Klein_Gordon_approximation_positive_negative_frequencies}) in the Klein-Gordon equation (\ref{Klein_Gordon}), we obtain
\begin{equation}\label{Schrodinger_free_mass_positive_negative_frequencies_error}
i \hbar \partial_0 g(\theta) = \left[ H_\text{M} + m c^2 \mathcal{O}(\epsilon^2) \right] g(\theta),
\end{equation}
which leads to the Schr\"{o}dinger equation (\ref{Schrodinger_free_mass_positive_negative_frequencies}).

From Eq.~(\ref{Schrodinger_free_mass_positive_negative_frequencies_error}) one can also derive the error associated to the approximation (\ref{Schrodinger_free_mass_positive_negative_frequencies}). The difference between the non-relativistic Hamiltonian $H_\text{M}$ and the exact fully-relativistic Hamiltonian $h_\text{KG}$ acting on non-relativistic states is of order
\begin{equation} \label{H_M_h_KG_error}
H_\text{M} - h_\text{KG} \sim \epsilon^2 mc^2.
\end{equation}

The equivalent of Eq.~(\ref{KG_scalar_product_f_nonrelativistic}) for $g(\theta)$ modes reads
\begin{equation}\label{KG_scalar_product_g_nonrelativistic}
( g(\theta), g(\theta') )_\text{KG} \approx  \frac{2 m}{\hbar^2} ( g(\theta), g(\theta') )_{L^2(\mathbb{R}^3)}, 
\end{equation}
which can be obtained by using Eq.~(\ref{positive_frequencies}) and the approximation (\ref{non_relativistic_limit}). The error associated to the approximation of the scalar products (\ref{KG_scalar_product_g_nonrelativistic}) comes directly from having replaced the time derivative of the modes with $m c^2/\hbar$ times such modes.  The relative error is, hence, of the order of $\epsilon$:
\begin{align} \label{scalar_product_error_order}
 \frac{( g(\theta), g(\theta') )_\text{KG}}{\frac{2 m}{\hbar^2} ( g(\theta), g(\theta') )_{L^2(\mathbb{R}^3)}} - 1  \sim \epsilon.
\end{align}

Equations (\ref{Schrodinger_free_mass_positive_negative_frequencies}) and (\ref{KG_scalar_product_g_nonrelativistic}) result again in the familiar description of free single-particle states in the position representation, as before. In this case, however, $g(\theta)$ represents a generic basis $|\theta \rangle $ for the single-particles space. The description of non-relativistic Fock states is given again by Eqs.~(\ref{Schrodinger_free_mass_n_particles}) and (\ref{KG_scalar_product_nonrelativistic}), with the definition of wave functions in a generic basis provided by Eq.~(\ref{wavefunction_g}).

Finally, we want to provide an analysis for an interacting scalar field. We work in the interaction picture. Therefore, the field $\hat{\phi}(t,\vec{x})$ is free --- i.e. solution of the Klein-Gordon equation (\ref{Klein_Gordon}) --- while any quantum state $| \phi(t) \rangle $ is time-evolved through an interacting potential $\hat{V}(t)$
\begin{equation} \label{Schrodinger_interaction}
i \hbar \partial_0 | \phi(t) \rangle = \hat{V}(t) | \phi(t) \rangle.
\end{equation}

In the interaction picture, the field $\hat{\phi}$ can still be expanded in terms of $g(\theta)$ and $h(\theta)$ modes as in Eq.~(\ref{free_field_positive_negative_frequencies}) and the Hilbert state can still be defined as the Fock space generated by the orthonormal free single-particle states $|\theta \rangle $.

We show that, in the non-relativistic limit, states, scalar product and Hamiltonian can be represented identically to the free case, with the only modifications coming from an extra term in the Hamiltonian. To see this, we use the modes $g(\theta)$ as representatives of $| \theta \rangle$, evolved with respect to the free theory.

A generic particles state $| \phi (t) \rangle$ is expanded with respect to the $| \bm{\theta}_n \rangle$ basis:
\begin{align}
| \phi (t) \rangle  = & \sum_{n=0}^\infty \sum_{\bm{\theta}_n} \tilde{\phi}_n (\bm{\theta}_n, t) | \bm{\theta}_n \rangle.
\end{align}
In this case, the $n$-particle wave function $\tilde{\phi}_n$ is time dependent, since the time evolution of $| \phi (t) \rangle$ in the interaction picture is given by Eq.~(\ref{Schrodinger_interaction}). This leads to a differential equation for $\tilde{\phi}_n$ that reads
\begin{equation} \label{Schrodinger_interaction_wavefunction}
i \hbar \partial_0 \tilde{\phi}_n (\bm{\theta}_n, t)  = \sum_{m=0}^\infty   \sum_{\bm{\theta}'_m}  \langle \bm{\theta}_n | \hat{V}(t) | \bm{\theta}'_m \rangle  \tilde{\phi}_m (\bm{\theta}'_m, t).
\end{equation}

The representative of the state $| \phi (t) \rangle$ in the Schr\"{o}dinger picture reads
\begin{equation}
\phi_n (t, \textbf{x}_n) = \left( \frac{2 m}{\hbar^2} \right)^{n/2} \sum_{\bm{\theta}_n} \tilde{\phi}_n (\bm{\theta}_n, t) \prod_{l=1}^n g(\theta_l,t,\vec{x}_l),
\end{equation}
where, differently from Eq.~(\ref{wavefunction_g}), $\tilde{\phi}_n $ is time dependent accordingly to Eq.~(\ref{Schrodinger_interaction_wavefunction}).

For interacting particles we still define non-relativistic states as the ones such that $\tilde{\phi}_n (\bm{\theta}_n, t) $ is non-vanishing only for non-relativistic frequencies $\omega(\theta)$. However, we also require potential energies that are very small with respect to the mass term. We therefore consider the following condition
\begin{equation}\label{non_relativistic_potential}
\langle \bm{\theta}_n | \hat{V}(t) | \bm{\theta}'_m \rangle  \sim \epsilon m c^2,
\end{equation}
so that Eq.~(\ref{Schrodinger_interaction_wavefunction}) is of order $\epsilon m c^2 \tilde{\phi}_n $.

Thanks to Eq.~(\ref{Schrodinger_interaction_wavefunction}) it is straightforward to prove that Eq.~(\ref{Schrodinger_free_mass_n_particles}) still holds, but with an additional potential term
\begin{align}\label{Schrodinger_interactive_mass_n_particles}
i \hbar \partial_0  \phi_n (t, \textbf{x}_n)  \approx & \sum_{l=1}^n \left( m c^2  - \frac{\hbar^2}{2 m} \nabla^2_{\vec{x}_l} \right) \phi (t, \textbf{x}_n)
 \nonumber \\
&+ \sum_{\bm{\theta}_n} \left( \frac{2 m}{\hbar^2} \right)^{n/2} \sum_{m=0}^\infty   \sum_{\bm{\theta}'_m} \langle \bm{\theta}_n | \hat{V}(t) | \bm{\theta}'_m \rangle   \nonumber \\
& \times \tilde{\phi}_m (\bm{\theta}'_m, t) \prod_{l=1}^n g(\theta_l,t,\vec{x}_l) .
\end{align}
Equation (\ref{Schrodinger_interactive_mass_n_particles}) can be identified as the NRQT Schr\"{o}dinger equation for particles with potential. It can be noticed that the error associated to Eq.~(\ref{Schrodinger_interactive_mass_n_particles}) is still of the order $\epsilon^2 mc^2$ [Eq.~(\ref{H_M_h_KG_error})], since the interacting part of Eq.~(\ref{Schrodinger_interactive_mass_n_particles}) has been exactly derived and the error associated to the time evolution only comes from the free part.

It is also possible to prove that Eq.~(\ref{KG_scalar_product_single_particle_nonrelativistic}) holds for non-relativistic interacting single-particles. Here, Eq.~(\ref{non_relativistic_potential}) plays an important role. Indeed, it suppresses the terms coming from the time derivative of $\tilde{\phi}_n $ [Eq.~(\ref{Schrodinger_interaction_wavefunction})] that appear as extra terms in Eq.~(\ref{KG_scalar_product_single_particle_nonrelativistic}). Moreover, the fact that $\hbar \partial_0 \tilde{\phi}_n$ is of order $\epsilon mc^2 \tilde{\phi}_n $ means that the relative error associated to the approximation (\ref{KG_scalar_product_single_particle_nonrelativistic}) is still of order $\epsilon$, as for the free case [Eq.~(\ref{scalar_product_error_order})].

The need for Eq.~(\ref{non_relativistic_potential}) implies that in order to have the same description of non-relativistic particles for free and interacting systems, we have to assume that the energy potential is small if compared to the mass term. The fact that the energy of the particles is close to their mass energy [Eq.~(\ref{non_relativistic_limit})] and that the potential energy is very small with respect to the mass [Eq.~(\ref{non_relativistic_potential})] means that also the kinetic energy of the particles is small. In this way we recover the definition of non-relativistic particles in terms of their velocity.

\subsection{Dirac field}\label{Minkowski_spacetime_Dirac_field}

In the previous section, we have been able to derive the familiar position representation of states, scalar product and Hamiltonian in the non-relativistic limit, starting from the fully relativistic description of scalar particles in QFT. A very similar result holds for Dirac fields $\hat{\psi}$.

Here, we show that non-relativistic Dirac particles can be described by wave functions, scalar product and Hamiltonian as prescribed by the NRQT. Specifically, the representation space of single-particles is $\mathbb{C}^2 \otimes L^2(\mathbb{R}^3) $ and the time evolution is given by a Schr\"{o}dinger equation similar to Eq.~(\ref{Schrodinger_interactive_mass_n_particles}). The difference with the scalar theory relies on the two spin degrees of freedom and the possibility to have interaction-spin coupling in the energy potential.

This section is organized as Sec.~\ref{Minkowski_spacetime_Scalar_field}. We start from the free theory and derive the NRQT description of non-relativistic particles with fixed momenta. We also show that the time evolution of these particles can be approximately described by the same Schr\"{o}dinger equation (\ref{Schrodinger_free_mass_n_particles}) of the scalar case. Then, we use a general decomposition of the field in positive and negative frequencies to derive the same $\mathbb{C}^2 \otimes L^2(\mathbb{R}^3)$ representation space, but with a general basis. Finally, we detail the interacting case and show that Eq.~(\ref{Schrodinger_interactive_mass_n_particles}) still holds, but with a potential operator that can generally break the spin degeneracy.

Here, we use the Dirac representation for the field $\hat{\psi}$ and its modes. Therefore, we identify $\hat{\psi}$ as a $4$ dimensional vector and any operator acting on the left as a $4 \times 4$ matrix.

Free Dirac fields in Minkowski spacetime are solutions of the Dirac equation, which reads
\begin{equation} \label{Dirac}
\left( i c \gamma^\mu \partial_\mu  - \frac{m c^2}{\hbar} \right) \hat{\psi} = 0, 
\end{equation}
where
\begin{align}\label{gamma_matrix}
& \gamma^0 = \frac{1}{c} \begin{pmatrix}
\mathbb{I} &0 \\
0 &-\mathbb{I}
\end{pmatrix}, & \gamma^i = \begin{pmatrix}
0 &\sigma^i \\
-\sigma^i &0
\end{pmatrix}
\end{align}
are gamma matrices, with $\mathbb{I}$ as $2 \times 2$ identity matrix and
\begin{align}
& \sigma^1 = \begin{pmatrix}
0 & 1 \\
1 & 0
\end{pmatrix}, & & \sigma^2 = \begin{pmatrix}
0 & -i \\
i & 0
\end{pmatrix}, & & \sigma^3 = \begin{pmatrix}
1 & 0 \\
0 & -1
\end{pmatrix}
\end{align}
as Pauli matrices. The anticommutation relation of gamma matrices is the following
\begin{equation}\label{gamma_matrices_anticommutating_rule}
\{ \gamma^\mu, \gamma^\nu \} = -2 \eta^{\mu\nu}.
\end{equation}
Moreover, $\gamma^0$ is defined to be hermitian, while $\gamma^i$ antihermitian
\begin{align}\label{gamma_matrices_hermitianity}
& (\gamma^0)^\dagger = \gamma^0, & (\gamma^i)^\dagger = -\gamma^i.
\end{align}

The usual decomposition of $\hat{\psi}$ in terms of modes with defined momenta and spin reads
\begin{equation} \label{free_Dirac_field}
\hat{\psi}(t,\vec{x}) = \sum_{s=1}^2 \int_{\mathbb{R}^3} d^3 k \left[ u_s(\vec{k},t,\vec{x}) \hat{c}_s(\vec{k}) + v_s(\vec{k},t,\vec{x}) \hat{d}_s^\dagger(\vec{k}) \right],
\end{equation}
where $\hat{c}_s(\vec{k})$ and $\hat{d}_s^\dagger(\vec{k})$ are, respectively, annihilation operators for particles and creation operators for antiparticles with momentum $k$ and spin number $s$, and have the following anticommutation relations
\begin{subequations}\label{free_Dirac_field_anticommutating_rules}
\begin{align}
& \{\hat{c}_s(\vec{k}),\hat{c}_{s'}(\vec{k'})\} = \{\hat{d}_s(\vec{k}),\hat{d}_{s'}(\vec{k'})\} = \{\hat{c}_s(\vec{k}),\hat{d}_{s'}(\vec{k'})\} = 0, \\
& \{\hat{c}_s(\vec{k}),\hat{c}_{s'}^\dagger(\vec{k'})\} = \{\hat{d}_s(\vec{k}),\hat{d}_{s'}^\dagger(\vec{k'})\} = \delta_{ss'} \delta^3(\vec{k}-\vec{k}').
\end{align}
\end{subequations}

The free Dirac modes $u_s(\vec{k},t,\vec{x})$ and $v_s(\vec{k},t,\vec{x})$ read
\begin{subequations}\label{free_Dirac_field_modes}
\begin{align}
u_s(\vec{k},t,\vec{x}) = &  \frac{c \gamma^0 \omega(k) - c \gamma^i k_i + mc^2/\hbar}{\sqrt{(2\pi)^3 2 \omega(k) [\omega(k)+mc^2/\hbar]}} e^{ -i\omega(k)t + i\vec{k} \cdot \vec{x} } \mathfrak{u}_s, \\
v_s(\vec{k},t,\vec{x}) = &  \frac{-c \gamma^0 \omega(k) + c \gamma^i k_i + mc^2/\hbar}{\sqrt{(2\pi)^3 2 \omega(k) [\omega(k)+mc^2/\hbar]}} e^{i\omega(k)t - i\vec{k} \cdot \vec{x}  } \mathfrak{v}_s,
\end{align}
\end{subequations}
with
\begin{align}\label{free_Dirac_field_basis}
\mathfrak{u}_1 = \begin{pmatrix}
1  \\
0 \\
0\\
0
\end{pmatrix},
& & \mathfrak{u}_2 =   \begin{pmatrix}
0  \\
1 \\
0\\
0
\end{pmatrix},
& & \mathfrak{v}_1 =  \begin{pmatrix}
0  \\
0 \\
1\\
0
\end{pmatrix},
& & \mathfrak{v}_2 =   \begin{pmatrix}
0  \\
0 \\
0\\
1
\end{pmatrix}.
\end{align}

$u_s(\vec{k})$ and $v_s(\vec{k})$ modes are orthonormal with respect to the $\mathbb{C}^4 \otimes L^2(\mathbb{R}^3) $ scalar product
\begin{equation}\label{Dirac_scalar_product}
( \psi, \psi' )_{\mathbb{C}^4 \otimes L^2(\mathbb{R}^3)} = \int_{\mathbb{R}^3} d^3x \psi^\dagger(t,\vec{x}) \psi'(t,\vec{x}),
\end{equation}
which is defined for any $t$ and any $\psi $, $ \psi'$ solutions of Eq.~(\ref{Dirac}). Indeed, it is possible to prove that 
\begin{subequations}\label{D_scalar_product_orthonormality_u_v}
\begin{align}
& ( u_s(\vec{k}), u_{s'}(\vec{k}') )_{\mathbb{C}^4 \otimes L^2(\mathbb{R}^3)} = \delta_{ss'} \delta^3(\vec{k}-\vec{k}'),\\
 & ( v_s(\vec{k}), v_{s'}(\vec{k}') )_{\mathbb{C}^4 \otimes L^2(\mathbb{R}^3)} = \delta_{ss'} \delta^3(\vec{k}-\vec{k}'),\\
  & ( u_s(\vec{k}), v_{s'}(\vec{k}') )_{\mathbb{C}^4 \otimes L^2(\mathbb{R}^3)} = 0.
\end{align}
\end{subequations}
Moreover, one can prove that Eq.~(\ref{Dirac_scalar_product}) is time independent for any $\psi $, $ \psi' $ solutions of Eq.~(\ref{Dirac}) thanks to Eqs.~(\ref{gamma_matrices_anticommutating_rule}), (\ref{gamma_matrices_hermitianity}) and an integration by parts:
\begin{align}
& \frac{d}{dt} ( \psi, \psi' )_{\mathbb{C}^4 \otimes L^2(\mathbb{R}^3)}\nonumber \\
 = & \frac{d}{dt} \int_{\mathbb{R}^3} d^3x \psi^\dagger \psi' \nonumber \\
 = & c^2 \frac{d}{dt} \int_{\mathbb{R}^3} d^3x \psi^\dagger\gamma^0 \gamma^0 \psi'  \nonumber \\
 = & c^2 \int_{\mathbb{R}^3} d^3x [ (\partial_0 \psi)^\dagger \gamma^0 \gamma^0 \psi'  + \psi^\dagger \gamma^0 \gamma^0 \partial_0 \psi' ]\nonumber \\
 = & c^2 \int_{\mathbb{R}^3} d^3x [(\gamma^0 \partial_0 \psi)^\dagger \gamma^0 \psi' + \psi^\dagger \gamma^0 \gamma^0 \partial_0 \psi' ]\nonumber \\
 = & c^2 \int_{\mathbb{R}^3} d^3x \left\lbrace \left[ \left( -\gamma^i \partial_i - i \frac{mc}{\hbar} \right) \psi\right]^\dagger \gamma^0 \psi' \right. \nonumber \\
 & \left. + \psi^\dagger \gamma^0 \left( -\gamma^i \partial_i - i \frac{mc}{\hbar} \right) \psi' \right\rbrace\nonumber \\
 = & c^2 \int_{\mathbb{R}^3} d^3x [ (\partial_i  \psi^\dagger) \gamma^i  \gamma^0 \psi'  - \psi^\dagger \gamma^0 \gamma^i \partial_i \psi'] \nonumber \\
 = & c^2 \int_{\mathbb{R}^3} d^3x [(\partial_i  \psi^\dagger) \gamma^i  \gamma^0 \psi' + \psi^\dagger \gamma^i \gamma^0 \partial_i \psi']\nonumber \\
 = & 0.
\end{align}

It can be noticed that the Dirac equation (\ref{Dirac}) is already put in a Schr\"{o}dinger equation form. Indeed, by acting on Eq.~(\ref{Dirac}) with a $\hbar c \gamma^0$ matrix and using Eq.~(\ref{gamma_matrices_anticommutating_rule}), one obtains
\begin{equation} \label{Dirac_schrodinger}
i \hbar \partial_0  \hat{\psi}  = h_\text{M}   \hat{\psi},
\end{equation}
with Hamiltonian
\begin{equation}\label{Dirac_Hamiltonian}
h_\text{M}  = - i \hbar c^2 \gamma^0 \gamma^i \partial_i + m c^3 \gamma^0.
\end{equation}
It can also be noticed that $h_\text{M}$ is hermitian with respect to the $\mathbb{C}^4 \otimes L^2(\mathbb{R}^3)$ scalar product:
\begin{equation}
( h_\text{M} \psi,  \psi' )_{\mathbb{C}^4 \otimes L^2(\mathbb{R}^3)}  = ( \psi, h_\text{M}  \psi' )_{\mathbb{C}^4 \otimes L^2(\mathbb{R}^3)}.
\end{equation}
This can be proven by using Eqs.~(\ref{gamma_matrices_anticommutating_rule}), (\ref{gamma_matrices_hermitianity}) and an integration by parts:
\begin{align}
& ( h_\text{M} \psi,  \psi' )_{\mathbb{C}^4 \otimes L^2(\mathbb{R}^3)} \nonumber \\
 = & \int_{\mathbb{R}^3} d^3x \left[ \left( - i \hbar c^2 \gamma^0 \gamma^i \partial_i + m c^3 \gamma^0 \right) \psi \right]^\dagger \psi'\nonumber \\
  = & \int_{\mathbb{R}^3} d^3x [ - i \hbar c^2 ( \partial_i \psi^\dagger) \gamma^i \gamma^0 + m c^3 \psi^\dagger \gamma^0] \psi'\nonumber \\
  = & \int_{\mathbb{R}^3} d^3x [  i \hbar c^2 ( \partial_i \psi^\dagger) \gamma^0 \gamma^i + m c^3 \psi^\dagger\gamma^0 ] \psi'\nonumber \\
  = & \int_{\mathbb{R}^3} d^3x \psi^\dagger   \left( - i \hbar c^2 \gamma^0 \gamma^i \partial_i + m c^3 \gamma^0 \right)\psi'\nonumber \\
    = & ( \psi, h_\text{M}  \psi' )_{\mathbb{C}^4 \otimes L^2(\mathbb{R}^3)}.
\end{align}

The quantum states $|s,\vec{k}\rangle = \hat{c}^\dagger_s (\vec{k}) | 0_\text{M} \rangle$ generate the Hilbert space of single-particles and they are orthonormal. This means that any state $| \psi \rangle$ can be decomposed in the Fock basis $| \textbf{s}_n,\textbf{k}_n \rangle$:
\begin{align}\label{free_Dirac_state_decomposition}
| \psi \rangle = & \sum_{n=0}^\infty \sum_{\textbf{s}_n} \int_{\mathbb{R}^{3n}} d^{3n} \textbf{k}_n \tilde{\psi}_n (\textbf{s}_n,\textbf{k}_n)| \textbf{s}_n,\textbf{k}_n \rangle,
\end{align}
with
\begin{equation}
(\textbf{s}_n,\textbf{k}_n) = ((s_1,\vec{k}_1), \dots, (s_n,\vec{k}_n))
\end{equation}
and with $\tilde{\psi}_n (\textbf{s}_n,\textbf{k}_n)$ antisymmetric with respect to spin-momenta variables. Equation (\ref{free_Dirac_state_decomposition}) is the equivalent of Eq.~(\ref{free_state_decomposition}) for Dirac particles and provides the definition of $\tilde{\psi}_n (\textbf{s}_n,\textbf{k}_n)$ as wave function for $| \psi \rangle$ in the spin-momentum representation.

Equivalently to Eq.~(\ref{free_wave_function}), the representative of the time-evolved state $| \psi (t) \rangle$ in the Schr\"{o}dinger picture reads
\begin{equation} \label{free_Dirac_wave_function}
\psi_n^{\bm{\alpha}_n} (t, \textbf{x}_n) = \sum_{\textbf{s}_n} \int_{\mathbb{R}^{3n}} d^{3n} \textbf{k}_n  \tilde{\psi}_n (\textbf{s}_n,\textbf{k}_n) \prod_{l=1}^n u^{\alpha_l}_{s_l}(\vec{k}_l, t, \vec{x}_l),
\end{equation}
where we have introduced the $4$ dimensional spinorial degrees of freedom through the $\alpha$ indexes:
\begin{equation}
\bm{\alpha}_n = \alpha_1 \dots \alpha_n.
\end{equation}
Single-particle wave functions, instead, can be written in the spinorial notation without indexes
\begin{equation} \label{free_Dirac_wave_function_single}
\psi_1 (t, \vec{x}) = \sum_{s=1}^2 \int_{\mathbb{R}^3} d^3 k  \tilde{\psi}_1 (s, \vec{k}) u_{s}(\vec{k}, t, \vec{x}).
\end{equation}

\begin{table}
\begin{tabular}{c|c|c|}
 & QFT & NRQT \\
\hline
$\langle \psi | \psi' \rangle$ & $(\psi_1, \psi'_1)_{\mathbb{C}^4 \otimes L^2(\mathbb{R}^3)}$ & $(\psi_1, \psi'_1)_{\mathbb{C}^2 \otimes L^2(\mathbb{R}^3)}$\\
\hline
Hamiltonian & $h_\text{M}$ & $H_\text{M}$ \\
\hline
\end{tabular}
\caption{Inner product (first line) and Hamiltonian (second line) for free Dirac single-particles. The left column is for the fully relativistic theory (QFT), while the right one is for the non-relativistic limit (NRQT).} \label{table_Dirac_Minkowski}
\end{table}

$\psi_1$ is solution of Eq.~(\ref{Dirac_schrodinger}) and, hence, its time evolution is provided by $h_\text{M}$. Moreover, the Hilbert product of any couple of single-particle states $| \psi \rangle $, $| \psi' \rangle $ can be written in terms of the $\mathbb{C}^4 \otimes L^2(\mathbb{R}^3)$ product of their wave functions
\begin{equation}\label{scalar_product_Dirac_single_particles}
\langle \psi | \psi' \rangle = (\psi_1, \psi'_1)_{\mathbb{C}^4 \otimes L^2(\mathbb{R}^3)}.
\end{equation}
This means that the single-particle content of the Dirac field can be fully described by spin-momentum wave functions (\ref{free_Dirac_wave_function_single}), $\mathbb{C}^4 \otimes L^2(\mathbb{R}^3)$ product and Hamiltonian $h_\text{M}$. This is summarized by Table \ref{table_Dirac_Minkowski} in its left column.

We may think that $\mathbb{C}^4 \otimes L^2(\mathbb{R}^3)$ is the representation space of the single-particle states. However, the orthonormal functions $u_s(\vec{k})$ do not provide a complete basis for $\mathbb{C}^4 \otimes L^2(\mathbb{R}^3)$, as it is possible to see from Eq.~(\ref{D_scalar_product_orthonormality_u_v}). The real representation space is actually a subspace of $\mathbb{C}^4 \otimes L^2(\mathbb{R}^3)$, namely the positive-frequency subspace of $\mathbb{C}^4 \otimes L^2(\mathbb{R}^3)$.

A generalization for an indefinite number of particles can be given by the wave functions (\ref{free_Dirac_wave_function}) and by the following Hilbert product
\begin{equation}
\langle \psi | \psi' \rangle = \sum_{n=0}^\infty (\psi_n, \psi'_n)_{\mathbb{C}^{4n} \otimes L^2(\mathbb{R}^{3n})},
\end{equation}
where
\begin{subequations}
\begin{align}
(\psi_n, \psi'_n)_{\mathbb{C}^{4n} \otimes L^2(\mathbb{R}^{3n})} = & \sum_{\bm{\alpha}_n} \int_{\mathbb{R}^{3n}} d^{3n} \textbf{x}_n [\psi^*_n (t, \textbf{x}_n)]^{\bm{\alpha}_n} \nonumber \\
& \times [\psi'_n (t, \textbf{x}_n)]^{\bm{\alpha}_n},
\end{align}
\begin{equation}
(\psi_0, \psi'_0)_{\mathbb{C}^0 \otimes L^2(\mathbb{R}^0)} = \psi_0^* \psi'_0.
\end{equation}
\end{subequations}

Once that fully-relativistic theory of Dirac particles has been provided, we move on the non-relativistic limit of the states. By taking the limit (\ref{dispersion_relation_approximation}), we obtain
\begin{equation}\label{Dirac_Minkowski_modes_nonrelativistic}
u_s(\vec{k},t,\vec{x}) \approx  \frac{1}{\sqrt{(2\pi)^3 }} \exp \left( -i\frac{m c^2 t}{\hbar} - i \frac{\hbar k^2 t}{2m} + i\vec{k} \cdot \vec{x} \right) \mathfrak{u}_s.
\end{equation}
From Eq.~(\ref{Dirac_Minkowski_modes_nonrelativistic}) it is immediate to see that the $u_s(\vec{k})$ modes cover the subspace of $\mathbb{C}^4 \otimes L^2(\mathbb{R}^3)$ with vanishing third and fourth spinorial components. More specifically, one can prove that
\begin{equation}\label{Dirac_representation_reduction_nonrelativistic_limit}
 \mathfrak{v}_{s'}^\dagger u_s(\vec{k}) \sim \epsilon \mathfrak{u}_{s''}^\dagger u_s(\vec{k}).
\end{equation}
This leads to a new representation for non-relativistic particle states, where the wave functions (\ref{free_Dirac_wave_function}) and the Hilbert product (\ref{Dirac_scalar_product}) can be considered with spinorial $\alpha$ indexes running through only the first two components. The representation space for non-relativistic particles can, hence, be identified with $\mathbb{C}^2 \otimes L^2(\mathbb{R}^3)$.

Moreover, the time evolution of $u_s(\vec{k})$ reads
\begin{align} \label{Schrodinger_Dirac_free_mass}
i \hbar \partial_0 u_s(\vec{k}) \approx H_\text{M}  u_s(\vec{k}),
\end{align}
which means that the spinorial components of $u_s(\vec{k})$ are approximately decoupled and are solutions of Eq.~(\ref{Schrodinger_free_mass}). It is also possible to notice that $H_\text{M}$ is hermitian with respect to the scalar product $( \psi, \psi' )_{\mathbb{C}^4 \otimes L^2(\mathbb{R}^3)} $:
\begin{equation}
( H_\text{M} \psi,  \psi' )_{\mathbb{C}^4 \otimes L^2(\mathbb{R}^3)}  = ( \psi, H_\text{M}  \psi' )_{\mathbb{C}^4 \otimes L^2(\mathbb{R}^3)},
\end{equation}
as it can be directly seen from the fact that
\begin{equation}\label{H_h_M}
H_\text{M} = \frac{h_\text{M}^2}{2 m c^2} + \frac{mc^2}{2},
\end{equation}
and, hence,
\begin{align}
& ( H_\text{M} \psi,  \psi' )_{\mathbb{C}^4 \otimes L^2(\mathbb{R}^3)}  \nonumber \\ = & \frac{1}{2 m c^2} ( h_\text{M} h_\text{M} \psi,  \psi' )_{\mathbb{C}^4 \otimes L^2(\mathbb{R}^3)} + \frac{mc^2}{2} ( \psi,  \psi' )_{\mathbb{C}^4 \otimes L^2(\mathbb{R}^3)} \nonumber \\
= & \frac{1}{2 m c^2} (  h_\text{M} \psi, h_\text{M} \psi' )_{\mathbb{C}^4 \otimes L^2(\mathbb{R}^3)} + \frac{mc^2}{2} ( \psi,  \psi' )_{\mathbb{C}^4 \otimes L^2(\mathbb{R}^3)} \nonumber \\
= & \frac{1}{2 m c^2} ( \psi,  h_\text{M} h_\text{M} \psi' )_{\mathbb{C}^4 \otimes L^2(\mathbb{R}^3)} + \frac{mc^2}{2} ( \psi,  \psi' )_{\mathbb{C}^4 \otimes L^2(\mathbb{R}^3)} \nonumber \\
= &( \psi, H_\text{M}  \psi' )_{\mathbb{C}^4 \otimes L^2(\mathbb{R}^3)}.
\end{align}
Equation (\ref{H_h_M}), on the other hand, can be derived from Eq.~(\ref{gamma_matrices_anticommutating_rule}) and the symmetry of second derivatives ($\partial_i \partial_j = \partial_j \partial_i$):
\begin{align}
\frac{h_\text{M}^2}{2 m c^2} + \frac{mc^2}{2} = &  - \frac{(\hbar c)^2}{2 m} \gamma^0 \gamma^i \gamma^0 \gamma^j \partial_i  \partial_j + \frac{m c^4}{2} \gamma^0 \gamma^0 \nonumber \\
&- i \frac{\hbar c^3}{2} (\gamma^0 \gamma^i \gamma^0 + \gamma^i \gamma^0 \gamma^0) \partial_i + \frac{mc^2}{2}\nonumber \\
 = &  \frac{(\hbar c)^2}{2 m} \gamma^0 \gamma^0 \gamma^i \gamma^j \partial_i  \partial_j  + mc^2 \nonumber \\
 = &  -\frac{\hbar^2}{2 m} \eta^{i j} \partial_i  \partial_j  + mc^2\nonumber \\
 = & H_\text{M}.
\end{align}

The result is that, in the non-relativistic limit, single particles are described as elements of $\mathbb{C}^2 \otimes L^2(\mathbb{R}^3)$, where $u_s^\alpha(\vec{k},t,\vec{x})$ is the wave function of a particle with momentum $\vec{k}$ and spin number $s$ and with spinorial index $\alpha$ running through the first two values. The states are also approximately evolved with respect to the Hamiltonian $H_\text{M}$. This description is listed in Table \ref{table_Dirac_Minkowski} on the right column and can be compared with the relativistic case, which is on the left column.

We have been able to derive the familiar description of Dirac particles in NRQT. General Fock states can be obtained from the singe-particle representation space $\mathbb{C}^2 \otimes L^2(\mathbb{R}^3)$ and from the following Schr\"{o}dinger equation
\begin{equation}\label{Schrodinger_Dirca_free_mass_n_particles}
i \hbar \partial_0  \psi_n^{\bm{\alpha}_n} \approx \sum_{l=1}^n \left( m c^2  - \frac{\hbar^2}{2 m} \nabla^2_{\vec{x}_l} \right) \psi_n^{\bm{\alpha}_n}.
\end{equation}

We want to provide the same description of non-relativistic particles but starting from generic real-frequency Dirac modes. For this reason, we use the general expression for $\hat{\psi}$ similar to Eq.~(\ref{free_field_positive_negative_frequencies})
\begin{equation}
\hat{\psi}(t,\vec{x}) = \sum_\theta \left[ u(\theta,t,\vec{x}) \hat{c}(\theta)  + v(\theta,t,\vec{x}) \hat{d}^\dagger(\theta) \right],
\end{equation}
with the difference that here the spin degrees of freedom introduces a further energy degeneracy and that the modes $u(\theta)$ and $v(\theta)$ have spinorial components. The time-dependency of $u(\theta)$ and $v(\theta)$ reads identically to Eq.~(\ref{positive_negative_frequencies})
\begin{subequations}\label{positive_negative_frequencies_Dirac}
\begin{align}
& u(\theta,t,\vec{x}) = \tilde{u}(\theta,\vec{x}) e^{-i \omega(\theta) t},\label{positive_frequencies_Dirac} \\
 & v(\theta,t,\vec{x}) = \tilde{v}(\theta,\vec{x}) e^{i \omega(\theta) t}.
\end{align}
\end{subequations}
Equation (\ref{positive_negative_frequencies_Dirac}) is guaranteed by the already-proven hermicity of $h_\text{M}$. $u(\theta)$ and $v(\theta)$ are also defined to be orthonormal with respect to the $\mathbb{C}^4 \otimes L^2(\mathbb{R}^3)$ product
\begin{subequations}\label{D_scalar_product_orthonormality_u_v_generic}
\begin{align}
& ( u(\theta), u(\theta') )_{\mathbb{C}^4 \otimes L^2(\mathbb{R}^3)}  =  \delta_{\theta\theta'} ,\\
& ( v(\theta), v(\theta') )_{\mathbb{C}^4 \otimes L^2(\mathbb{R}^3)}  =  \delta_{\theta\theta'}, \\
& ( u(\theta), v(\theta') )_{\mathbb{C}^4 \otimes L^2(\mathbb{R}^3)} = 0.
\end{align}
\end{subequations}

Any Fock state $| \psi \rangle$ is expanded with respect to the single-particle basis $| \theta \rangle =  \hat{c}^\dagger(\theta) | 0_\text{M} \rangle$ as in Eq.~(\ref{general_Fock_expansion})
\begin{align} \label{general_Fock_Dirac_expansion}
| \psi \rangle  = & \sum_{n=0}^\infty \sum_{\bm{\theta}_n} \tilde{\psi}_n (\bm{\theta}_n) | \bm{\theta}_n \rangle.
\end{align}
The representative of state $| \psi (t) \rangle$ in the Schr\"{o}dinger picture reads similarly to Eq.~(\ref{free_Dirac_wave_function})
\begin{equation}\label{Dirac_wavefunction_general}
\psi^{\bm{\alpha}_n}_n (t, \textbf{x}_n) =  \sum_{\bm{\theta}_n} \tilde{\psi}_n (\bm{\theta}_n) \prod_{l=1}^n u^{\alpha_l}(\theta_l, t, \vec{x}_l).
\end{equation}
It is straightforward to prove Eq.~(\ref{scalar_product_Dirac_single_particles}) for single particles with the new definition of $\psi_1 (t, \vec{x})$ given by Eq.~(\ref{Dirac_wavefunction_general}).

The result is the description of single particles through the $\mathbb{C}^4 \otimes L^2(\mathbb{R}^3)$ Hilbert space. The new basis is identified by particles with quantum numbers $\theta$. Thanks to Eq.~(\ref{D_scalar_product_orthonormality_u_v_generic}) we notice again that the representation space is actually a subspace of $\mathbb{C}^4 \otimes L^2(\mathbb{R}^3)$.

For non-relativistic particles, we can identify such subspace as the one in which the third and fourth spinorial components are always vanishing. Indeed, it is possible to prove the equivalent of Eq.~(\ref{Dirac_representation_reduction_nonrelativistic_limit}) for the $u(\theta)$ modes
\begin{equation}\label{Dirac_representation_reduction_nonrelativistic_limit_general}
 \mathfrak{v}_s^\dagger u(\theta) \sim \epsilon \mathfrak{u}_{s'}^\dagger u(\theta).
\end{equation}
The proof of Eq.~(\ref{Dirac_representation_reduction_nonrelativistic_limit_general}) follows from the fact that $u(\theta)$ is solution of the Dirac equation (\ref{Dirac}) and, hence,
\begin{equation} \label{Dirac_representation_reduction_nonrelativistic_limit_general_Dirac}
\mathfrak{u}_s^\dagger \left[ c \gamma^0 \omega(\theta) + i c \gamma^i \partial_i  - \frac{m c^2}{\hbar} \right] u(\theta) = 0.
\end{equation}
Equation (\ref{Dirac_representation_reduction_nonrelativistic_limit_general_Dirac}) can be put in the following form
\begin{equation} \label{Dirac_representation_reduction_nonrelativistic_limit_general_Dirac_2}
\frac{mc^2}{\hbar} \epsilon \mathfrak{u}_s^\dagger  u(\theta) + i c \sum_{s'=1}^2 \mathfrak{u}_s^\dagger \gamma^i \mathfrak{v}_{s'}  \mathfrak{v}_{s'}^\dagger \partial_i   u(\theta) = 0,
\end{equation}
which leads to Eq.~(\ref{Dirac_representation_reduction_nonrelativistic_limit_general}).

One can prove Eq.~(\ref{Schrodinger_Dirac_free_mass}) for the mode $u(\theta)$ in the following way. It is known that the components of any solution of the Dirac equation are also solution of the Klein-Gordon equation (\ref{Klein_Gordon}) with the same mass. This fact can be proven by multiplying Eq.~(\ref{Dirac}) with $ i c \gamma^\mu \partial_\mu + m c^2/\hbar$ on the left and exploiting the anticommutation relation (\ref{gamma_matrices_anticommutating_rule}). This means that the $u(\theta)$ modes are also solutions of the Klein-Gordon equation 
\begin{equation}\label{Klein_Gordon_Dirac}
\left[ c^2 \eta^{\mu\nu} \partial_\mu \partial_\nu - \left( \frac{mc^2}{\hbar} \right)^2 \right] u(\theta) = 0.
\end{equation}
We can, at this point, use the same arguments of Sec.~\ref{Minkowski_spacetime_Scalar_field} that have led to Eq.~(\ref{Schrodinger_free_mass_positive_negative_frequencies}) in order to prove that
\begin{equation}\label{Schrodinger_Dirac_free_mass_positive_negative_frequencies}
i \hbar \partial_0 u(\theta) \approx H_\text{M}  u(\theta).
\end{equation}

The error associated to the approximation (\ref{Schrodinger_Dirac_free_mass_positive_negative_frequencies}) can be identified with the equivalent of Eq.~(\ref{H_M_h_KG_error}) for Dirac fields
\begin{equation} \label{H_M_h_M_error}
H_\text{M} - h_\text{M} \sim \epsilon^2 mc^2,
\end{equation}
which, in turn, can be obtained from Eq.~(\ref{Schrodinger_free_mass_positive_negative_frequencies_error}) for $u(\theta)$ modes. From Eqs.~(\ref{H_M_h_KG_error}) and (\ref{H_M_h_M_error}) one can derive the error made by considering scalar and Dirac states identical in their time evolution [Eqs.~(\ref{Schrodinger_free_mass_positive_negative_frequencies}) and (\ref{Schrodinger_Dirac_free_mass_positive_negative_frequencies})]:
\begin{equation} \label{h_KG_h_M_error}
h_\text{KG} - h_\text{M} \sim \epsilon^2 mc^2.
\end{equation}
Equation (\ref{h_KG_h_M_error}) implies that corrective terms of Eqs.~(\ref{Schrodinger_free_mass_positive_negative_frequencies}) and (\ref{Schrodinger_Dirac_free_mass_positive_negative_frequencies}) that spoil the difference between scalar and Dirac fields in the Minkowski spacetime can be found at order $\epsilon^2$.

A second error associated to the non-relativistic limit comes from considering the third and the fourth spinorial component of $u(\theta)$ vanishing quantities. Such approximation allowed us to replace the exact $\mathbb{C}^4 \otimes L^2(\mathbb{R}^3)$ scalar product with the $\mathbb{C}^2 \otimes L^2(\mathbb{R}^3)$ scalar product. The relative error can be obtained from Eq.~(\ref{Dirac_representation_reduction_nonrelativistic_limit_general}) and is of order $\epsilon$, as in the scalar case [Eq.~(\ref{scalar_product_error_order})].

Finally, we consider an interacting Dirac field and by following the same steps of Sec.~\ref{Minkowski_spacetime_Scalar_field} we see that interacting particles can be still described in the representation space of free particles. The difference between the interacting and the free theory is only given by the presence of a potential energy in the approximated Schr\"{o}dinger equation. Such term can introduce spin interactions that cannot appear in the scalar theory.

The representative of any state $| \psi(t) \rangle$ in the Schr\"{o}dinger picture reads
\begin{equation}\label{Dirac_wavefunction_general_interaction}
\psi^{\bm{\alpha}_n}_n (t, \textbf{x}_n) =  \sum_{\bm{\theta}_n} \tilde{\psi}_n (\bm{\theta}_n, t)  \prod_{l=1}^n u^{\alpha_l}(\theta_l, t, \vec{x}_l),
\end{equation}
where, differently from Eq.~(\ref{Dirac_wavefunction_general}), $\tilde{\psi}_n (\bm{\theta}_n, t)$ is time dependent. The time evolution of Eq.~(\ref{Dirac_wavefunction_general_interaction}) in the non-relativistic limit reads
\begin{align}\label{Schrodinger_Dirca_free_mass_n_particles_interaction}
i \hbar \partial_0  \psi_n^{\bm{\alpha}_n} (t, \textbf{x}_n) \approx & \sum_{l=1}^n \left( m c^2  - \frac{\hbar^2}{2 m} \nabla^2_{\vec{x}_l} \right) \psi_n^{\bm{\alpha}_n} (t, \textbf{x}_n)  \nonumber \\
&+ \sum_{\bm{\theta}_n}  \sum_{m=0}^\infty   \sum_{\bm{\theta}'_m} \langle \bm{\theta}_n | \hat{V}(t) | \bm{\theta}'_m \rangle  \nonumber \\
& \times \tilde{\psi}_m (\bm{\theta}'_m, t)  \prod_{l=1}^n u^{\alpha_l}(\theta_l, t, \vec{x}_l),
\end{align}
where, in this case, $ \langle \bm{\theta}_n | \hat{V}(t) | \bm{\theta}'_m \rangle$ are the matrix elements of a potential $\hat{V}(t)$ that comes from a Dirac interacting Lagrangian. The quantum numbers $\theta$ also contain spinorial degrees of freedom, and, hence, $\langle \bm{\theta}_n | \hat{V}(t) | \bm{\theta}'_m \rangle$ can break the spin degeneracy present in the free theory. As explained in Sec.~\ref{Minkowski_spacetime_Scalar_field}, we obtain Eq.~(\ref{Schrodinger_Dirca_free_mass_n_particles_interaction}) by supplementing the definition of non-relativistic limit of the free theory (\ref{non_relativistic_limit}) with Eq.~(\ref{non_relativistic_potential}).

\section{Non-Minkowski spacetime}\label{NonMinkowksi_spacetime}

At variance with Sec.~\ref{Minkowski_spacetime}, here we work with coordinates $(T,\vec{X})$ and metric $g_{\mu\nu}(T,\vec{X})$ which represent a curved spacetime. The aim of this section is to derive a description for non-relativistic states of scalar $\hat{\Phi}(T,\vec{X})$ and Dirac $\hat{\Psi}(T,\vec{X})$ field.

We start from the description of fully-relativistic particles states in static spacetimes --- see for instance \cite{Wald:1995yp} --- and then we perform the non-relativistic limit. We show how the representation of non-relativistic states changes from the Minkowski to the non-Minkowski case. We also derive the Schr\"{o}dinger equation for particles affected by the curvature and the consequent precision needed to distinguish between scalar and Dirac fields.

\subsection{Scalar field}\label{NonMinkowksi_spacetime_Scalar_field}

The field considered in the present section is scalar. As in Sec.~\ref{Minkowski_spacetime_Scalar_field}, we start by reviewing the relativistic theory of particles for the free scalar field $\hat{\Phi}$. Each positive-frequency mode is associated to a single-particle state and the Klein-Gordon product in curved spacetime is used as Hilbert product for the single-particle space. Then we perform the non-relativistic limit and show that such product can be approximated by the $L^2(\mathbb{R}^3)$ inner product with a metric-dependent measure. Moreover, we show that the quantum states are solutions of a metric-dependent Schr\"{o}dinger equation. Finally, we extend the theory to the interacting case by introducing a potential energy in the Schr\"{o}dinger equation.

We consider a free scalar field $\hat{\Phi}$ that is solution of the Klein-Gordon equation in curved spacetime
\begin{equation} \label{Klein_Gordon_curved}
\left[ \frac{c^2}{\sqrt{-g}} \partial_\mu \left( \sqrt{-g} g^{\mu \nu}\partial_\nu   \right) - \left( \frac{mc^2}{\hbar}\right)^2 \right] \hat{\Phi} = 0,
\end{equation}
where $g$ is the determinant of $g_{\mu\nu}$. We also consider the curved Klein-Gordon scalar product
\begin{align}\label{KG_curved_scalar_product}
& ( \Phi, \Phi' )_\text{CKG} =  -\frac{i}{\hbar c} \int_{\mathbb{R}^3} d^3X  \sqrt{-g(T,\vec{X})} g^{0 \mu}(T,\vec{X}) \nonumber \\
& \times \left[ \Phi^*(T,\vec{X}) \partial_\mu \Phi'(T,\vec{X})  -  \Phi'(T,\vec{X}) \partial_\mu \Phi^*(T,\vec{X}) \right].
\end{align}

$( \Phi, \Phi' )_\text{CKG} $ is time independent for solutions of the curved Klein-Gordon equation (\ref{Klein_Gordon_curved}). This can be proven by using the integration by parts
\begin{align}
& \frac{d}{dT}( \Phi, \Phi' )_\text{CKG} \nonumber \\ 
 = & -\frac{i}{\hbar c} \int_{\mathbb{R}^3} d^3X \left[ (\partial_0 \Phi^*) \sqrt{-g} g^{0 \mu}  \partial_\mu \Phi'  \right. \nonumber \\
& + \Phi^* \partial_0 (\sqrt{-g} g^{0 \mu}  \partial_\mu \Phi') - (\partial_0 \Phi') \sqrt{-g} g^{0 \mu}  \partial_\mu \Phi^* \nonumber \\
& \left. -  \Phi' \partial_0 ( \sqrt{-g} g^{0 \mu}  \partial_\mu \Phi^*) \right]\nonumber \\
 = & -\frac{i}{\hbar c} \int_{\mathbb{R}^3} d^3X \left\lbrace (\partial_0 \Phi^*) \sqrt{-g} g^{0 \mu}  \partial_\mu \Phi' \right. \nonumber \\
& + \Phi^* \left[ -\partial_i (\sqrt{-g} g^{i \mu}  \partial_\mu) + \sqrt{-g} \left( \frac{mc}{\hbar} \right)^2 \right] \Phi'  \nonumber \\
& - (\partial_0 \Phi') \sqrt{-g} g^{0 \mu}  \partial_\mu \Phi^* \nonumber \\
& \left. -  \Phi'  \left[ -\partial_i (\sqrt{-g} g^{i \mu}  \partial_\mu) + \sqrt{-g} \left( \frac{mc}{\hbar} \right)^2 \right]  \Phi^* \right\rbrace\nonumber \\
 = & -\frac{i}{\hbar c} \int_{\mathbb{R}^3} d^3X \left[ (\partial_0 \Phi^*) \sqrt{-g} g^{0 \mu}  \partial_\mu \Phi' \right. \nonumber \\
& + (\partial_i \Phi^* ) \sqrt{-g} g^{i \mu}  \partial_\mu \Phi' - (\partial_0 \Phi') \sqrt{-g} g^{0 \mu}  \partial_\mu \Phi^*  \nonumber \\
& \left.  -   (\partial_i \Phi') \sqrt{-g} g^{i \mu}  \partial_\mu   \Phi^* \right]\nonumber \\
 = & -\frac{i}{\hbar c} \int_{\mathbb{R}^3} d^3X \left[ (\partial_\nu \Phi^*) \sqrt{-g} g^{\nu \mu}  \partial_\mu \Phi' \right. \nonumber \\
& \left.  -   (\partial_\nu \Phi') \sqrt{-g} g^{\nu \mu}  \partial_\mu   \Phi^* \right]\nonumber \\
 = & 0.
\end{align}
For this reason $( \Phi, \Phi' )_\text{CKG}$ can be used as Hilbert product for positive-frequency modes.

By expanding $\hat{\Phi}$ in terms of modes with real frequencies with respect to the time $T$, we obtain
\begin{equation}\label{Klein_Gordon_curved_Phi}
\hat{\Phi}(T,\vec{X}) = \sum_\theta\left[ G(\theta,T,\vec{X}) \hat{A}(\theta)  + H(\theta,T,\vec{X}) \hat{B}^\dagger(\theta) \right],
\end{equation}
where $\hat{A}(\theta)$ ($\hat{B}(\theta)$) is the annihilation operator associated to the particle (antiparticle) mode $G(\theta)$ ($H^*(\theta)$).

The $G(\theta)$ and $H(\theta)$ modes are defined to be orthonormal with respect to the curved Klein-Gordon scalar product (\ref{KG_curved_scalar_product}):
\begin{subequations}\label{KG_scalar_curved_product_orthonormality_G}
\begin{align}
& ( G(\theta), G(\theta'))_\text{CKG} = \delta_{\theta\theta'} , \\
 & ( H(\theta), H(\theta'))_\text{CKG}  = - \delta_{\theta\theta'} ,\\
 & ( G(\theta), H(\theta'))_\text{CKG} = 0.
\end{align}
\end{subequations}
As in Eq.~(\ref{positive_negative_frequencies}), the definition of positive and negative frequency modes is expressed by
\begin{subequations}\label{scalar_curved_modes_tilde}
\begin{align} 
& G(\theta,T,\vec{X}) = \tilde{G}(\theta,\vec{X}) e^{-i \Omega(\theta) T}, \\
 & H(\theta,T,\vec{X}) = \tilde{H}(\theta,\vec{X}) e^{i \Omega(\theta) T}.
\end{align}
\end{subequations}

It is important to mention that the expansion of $\hat{\Phi}$ in positive and negative frequency modes is not always possible. For some metrics, the ansatz (\ref{scalar_curved_modes_tilde}) is not compatible with Eq.~(\ref{Klein_Gordon_curved}). Condition for the validity of Eq.~(\ref{scalar_curved_modes_tilde}) is given by a static spacetime
\begin{align}\label{metric_constraint_static}
& \partial_0 g^{\mu \nu} = 0, & g^{0i} = g^{i0} = 0.
\end{align}
Indeed, Eq.~(\ref{Klein_Gordon_curved}) for $G(\theta) $ becomes a Schr\"{o}dinger equation with eigenvalues proportional to $\Omega^2$
\begin{equation}\label{scalar_curved_Schrodinger_equation_tilde}
 (\hbar \Omega)^2 G(\theta) = H_\text{CKG} G(\theta), 
\end{equation}
with Hamiltonian
\begin{align}\label{Schrodinger_equation_tilde_Hamiltonian}
H_\text{CKG} = & g_{0 0} \left[  \frac{\hbar^2}{ \sqrt{-g}} \partial_i \left( \sqrt{-g} g^{i j} \partial_j \right)  - (mc)^2 \right],
\end{align}
that is positive with respect to the curved Klein-Gordon scalar product (\ref{KG_curved_scalar_product}) for positive-frequency modes. The positivity of $H_\text{CKG}$ guarantees the existence of real $\Omega$ for Eq.~(\ref{scalar_curved_Schrodinger_equation_tilde}).

It is possible to prove that $H_\text{CKG}$ is positive thanks to the following identity
\begin{align}\label{scalar_curved_Klein_Gordon_hamiltonian_positivity}
( \Phi, H_\text{CKG} \Phi' )_\text{CKG} = & \delta^{i j} ( H_i \Phi, H_j \Phi' )_\text{CKG} \nonumber \\
& + ( H_0 \Phi, H_0 \Phi' )_\text{CKG},
\end{align}
with
\begin{align}
& H_0 =  m c^2 e^0{}_0 , & H_i = \hbar c e^0{}_0 e_i{}^j \partial_j,
\end{align}
where $e_\alpha{}^\mu$ is the vierbein field defined as
\begin{equation}
e_\alpha{}^\mu e_\beta{}^\nu g_{\mu \nu} =  \eta_{\alpha \beta}
\end{equation}
and with $e^\alpha{}_\mu$ as inverse. Equation (\ref{scalar_curved_Klein_Gordon_hamiltonian_positivity}), in turn, can be proven by using the static spacetime condition (\ref{metric_constraint_static}), which in terms of the vierbein field reads
\begin{align}\label{vierbein_constraint_static}
& \partial_0 e_\alpha{}^\mu = 0, & e_i{}^0 = e_0{}^i = 0.
\end{align}
The product (\ref{KG_curved_scalar_product}) in static spacetimes reads
\begin{align}\label{KG_curved_scalar_product_2}
& ( \Phi, \Phi' )_\text{CKG} = -\frac{i}{\hbar c} \int_{\mathbb{R}^3} d^3X  \sqrt{-g(\vec{X})} g^{0 0}(\vec{X}) \nonumber \\
& \times \left[ \Phi^*(T,\vec{X}) \partial_0 \Phi'(T,\vec{X})  -  \Phi'(T,\vec{X}) \partial_0 \Phi^*(T,\vec{X}) \right].
\end{align}
By integrating by parts, one obtains
\begin{align}\label{KG_curved_scalar_product_2_proof}
&  ( \Phi, H_\text{CKG} \Phi' )_\text{CKG}\nonumber \\ = & -i\frac{\hbar}{c} \int_{\mathbb{R}^3} d^3X  \{ \Phi^* [\partial_0 \partial_i \left( \sqrt{-g} g^{i j} \partial_j \right) \Phi'] \nonumber \\
& -  [\partial_i \left( \sqrt{-g} g^{i j} \partial_j \right) \Phi'] \partial_0 \Phi^*  \} \nonumber \\
& + i \frac{m^2 c}{\hbar} \int_{\mathbb{R}^3} d^3X \sqrt{-g}  \left( \Phi^* \partial_0 \Phi' -  \Phi' \partial_0 \Phi^*  \right) \nonumber \\
  = & -i\frac{\hbar}{c} \int_{\mathbb{R}^3} d^3X  \sqrt{-g} g^{i j} [ - (\partial_i \Phi^*) \partial_0 \partial_j \Phi' +  (\partial_j \Phi') \partial_i \partial_0 \Phi^*  ] \nonumber \\
&+ i \frac{m^2 c}{\hbar} \int_{\mathbb{R}^3} d^3X  \sqrt{-g}  \left( \Phi^* \partial_0 \Phi' -  \Phi' \partial_0 \Phi^*  \right) \nonumber \\
  = &  i \hbar c \int_{\mathbb{R}^3} d^3X   \sqrt{-g} g^{0 0}  e^0{}_0 e^0{}_0 \eta^{i j} e_i{}^{i'} e_j{}^{j'} [ - (\partial_{i'} \Phi^*)  \partial_0 \partial_{j'} \Phi'\nonumber \\
& + ( \partial_{j'} \Phi') \partial_{i'} \partial_0 \Phi^*  ]  \nonumber \\
& - i \frac{m^2c^3}{\hbar} \int_{\mathbb{R}^3} d^3X   \sqrt{-g} g^{0 0} e^0{}_0 e^0{}_0 \left( \Phi^* \partial_0 \Phi' - \Phi' \partial_0 \Phi^*  \right)\nonumber \\
=  & \delta^{i j} ( H_i \Phi, H_j \Phi' )_\text{CKG} + ( H_0 \Phi, H_0 \Phi' )_\text{CKG},
\end{align}
which proves Eq. (\ref{scalar_curved_Klein_Gordon_hamiltonian_positivity}).

Seemingly, one can prove that $H_\text{CKG}$ is positive with respect to the following scalar product
\begin{align}\label{KG_curved_scalar_product_2_nonrelativistic}
( \Phi, \Phi' )_{L^2_S(\mathbb{R}^3)} = & - c \int_{\mathbb{R}^3} d^3X  \sqrt{-g(\vec{X})} g^{0 0}(\vec{X}) \nonumber \\
& \times \Phi^*(\vec{X}) \Phi'(\vec{X}),
\end{align}
which can be seen as the $L^2(\mathbb{R}^3)$ inner product with a metric-dependent measure. The positivity of $H_\text{CKG}$ with respect to such product can still be obtained from an identity similar to Eq.~(\ref{scalar_curved_Klein_Gordon_hamiltonian_positivity}):
\begin{align}
( \Phi, H_\text{CKG} \Phi' )_{L^2_S(\mathbb{R}^3)} = & \delta^{i j} ( H_i \Phi, H_j \Phi' )_{L^2_S(\mathbb{R}^3)} \nonumber \\
& + ( H_0 \Phi, H_0 \Phi' )_{L^2_S(\mathbb{R}^3)}.
\end{align}
This scalar product will appear in the non-relativistic limit and can be interpreted as the non-Minkowski version of the usual $L^2(\mathbb{R}^3)$ inner product.

As in Sec.~\ref{Minkowski_spacetime_Scalar_field}, we interpret the Klein-Gordon equation for positive-frequency solutions as a Schr\"{o}dinger equation
\begin{equation}
i \hbar \partial_0 G(\theta) = h_\text{CKG} G(\theta),
\end{equation}
with Hamiltonian $h_\text{CKG}$ that is the square root of $H_\text{CKG}$. The equivalent of Eq.~(\ref{h_KG_H_KG}) in curved spacetime reads
\begin{equation}\label{h_cKG_H_cKG}
h_\text{CKG} = \sqrt{H_\text{CKG}}.
\end{equation}
We will show that such Hamiltonian can be approximated by a free single-particle Hamiltonian modified by the curvature.

\begin{table}
\begin{tabular}{c|c|c|}
 & QFTCS & NRQTCS \\
\hline
$\langle \Phi | \Phi' \rangle $ & $[\hbar^2/(2m)] ( \Phi_1, \Phi'_1 )_\text{CKG}$ & $ ( \Phi_1 , \Phi'_1 )_{L^2_S(\mathbb{R}^3)}$\\
\hline
Hamiltonian & $h_\text{CKG}$ & $H_\text{S}$ \\
\hline
\end{tabular}
\caption{Inner product (first line) and Hamiltonian (second line) for free scalar single-particles in curved spacetime. The left column is for the fully relativistic theory (QFTCS), while the right one is for the non-relativistic limit (NRQTCS).} \label{table_scalar_nonMinkowski}
\end{table}

In summary, the fully-relativistic single-particle description of the field is defined by the Hamiltonian $h_\text{CKG}$ and the scalar product $( \Phi, \Phi' )_\text{CKG}$, as shown by the left column of Table \ref{table_scalar_nonMinkowski}. Instead, general Fock states $| \Phi \rangle$ are represented in the Schr\"{o}dinger picture by
\begin{equation}\label{wavefunction_g_curved}
\Phi_n (T, \textbf{X}_n) =  \left( \frac{2 m}{\hbar^2} \right)^{n/2} \sum_{\bm{\theta}_n} \tilde{\Phi}_n (\bm{\theta}_n)    \prod_{l=1}^n G(\theta_l,T,\vec{X}_l),
\end{equation}
where $ \tilde{\Phi}_n (\bm{\theta}_n)$ is defined from the decomposition of $| \Phi \rangle$ in the Fock space, similarly to Eq.~(\ref{general_Fock_expansion})
\begin{equation} \label{general_Fock_expansion_curved}
| \Phi \rangle  = \sum_{n=0}^\infty \sum_{\bm{\theta}_n} \tilde{\Phi}_n (\bm{\theta}_n) | \bm{\theta}_n \rangle
\end{equation}
and is symmetric with respect to $\theta_1, \dots, \theta_n$.

In the non-Minkowski spacetime, we still refer to the non-relativistic limit as 
\begin{equation}\label{non_relativistic_limit_curved}
\left| \frac{\hbar \Omega}{mc^2} - 1 \right| \ll 1.
\end{equation}
We want to show that $G(\theta)$ is approximately solution to a Schr\"{o}dinger equation
\begin{equation} \label{Schrodinger_curved}
i \hbar \partial_0 G(\theta) \approx H_\text{S} G(\theta),
\end{equation}
with Hamiltonian
\begin{equation}\label{H_nM}
H_\text{S} = \frac{m c^2}{2} \left( 1 - \frac{g_{00}}{c^2} \right) + \frac{\hbar^2 g_{00}}{2 m c^2 \sqrt{-g}} \partial_i \left( \sqrt{-g} g^{i j}\partial_j \right)
\end{equation}
and that the curved Klein-Gordon scalar product $( \Phi, \Phi' )_\text{CKG}$ is approximated by $( \Phi, \Phi' )_{L^2_S(\mathbb{R}^3)}$
\begin{equation}\label{cKG_scalar_curved_product_G_nonrelativistic}
( G(\theta), G(\theta') )_\text{CKG}  \approx \frac{2 m}{\hbar^2} ( G(\theta), G(\theta') )_{L^2_S(\mathbb{R}^3)}.
\end{equation}
In this way, we show that the non-relativistic single-particle description of the field is defined by the Hamiltonian $H_\text{S}$ and the scalar product $( \Phi, \Phi' )_{L^2_S(\mathbb{R}^3)}$. The result can be seen as the equivalent of Eqs.~(\ref{Schrodinger_free_mass_positive_negative_frequencies}) and (\ref{KG_scalar_product_g_nonrelativistic}) in curved spacetime and are summarized by the right column of Table \ref{table_scalar_nonMinkowski}.

$H_\text{S}$ is hermitian with respect to the curved Klein-Gordon scalar product (\ref{KG_curved_scalar_product_2}) and the product given by Eq.~(\ref{KG_curved_scalar_product_2_nonrelativistic}), since it can also be written as
\begin{equation}\label{H_nM_H_cKG}
H_\text{S} = \frac{H_\text{CKG}}{2mc^2} + \frac{m c^2}{2}
\end{equation}
and $H_\text{CKG}$ is hermitian with respect to both products.

The non-relativistic description of states with indefinite numbers of particles is given by the wave functions $\Phi_n$ of Eq.~(\ref{wavefunction_g_curved}), the Fock extension of the $( \Phi, \Phi' )_{L^2_S(\mathbb{R}^3)}$ scalar product
\begin{equation}\label{KG_scalar_product_nonrelativistic_curved}
\langle \Phi | \Phi' \rangle \approx \sum_{n=0}^\infty ( \Phi_n , \Phi'_n )_{L^2_S(\mathbb{R}^{3n})},
\end{equation}
and the following Schr\"{o}dinger equation
\begin{align}\label{Schrodinger_free_mass_n_particles_curved}
i \hbar \partial_0  \Phi_n (T, \textbf{X}_n) \approx & \sum_{l=1}^n \left\lbrace \frac{m c^2}{2} \left[ 1 - \frac{g_{00}(\vec{X}_l)}{c^2} \right] \right. \nonumber \\ & \left. + \frac{\hbar^2 g_{00}(\vec{X}_l)}{2 m c^2} \nabla^2_{\vec{X}_l} \right\rbrace \Phi_n (T, \textbf{X}_n),
\end{align}
where, in this case,
\begin{subequations}
\begin{align}
& ( \Phi_n, \Phi'_n )_{L^2_S(\mathbb{R}^{3n})} = (- c)^n \int_{\mathbb{R}^{3n}} d^{3n}\textbf{X}_n\nonumber \\
& \times \left[ \prod_{l=1}^n \sqrt{-g(\vec{X}_l)} g^{0 0}(\vec{X}_l) \right]  \Phi^*_n(\textbf{X}_n) \Phi'_n(\textbf{X}_n),
\end{align}
\begin{equation}
( \Phi_0 , \Phi'_0 )_{L^2_S(\mathbb{R}^0)} = \Phi_0^* \Phi'_0.
\end{equation}
\end{subequations}
and
\begin{equation}
\nabla^2_{\vec{X}} = \frac{1}{\sqrt{-g(\vec{X})}} \frac{\partial}{\partial X^i} \left[ \sqrt{-g(\vec{X})} g^{i j}(\vec{X})\frac{\partial}{\partial X^j} \right].
\end{equation}

A way to approximate Eq.~(\ref{Klein_Gordon_curved}) as a Schr\"{o}dinger equation is to replace the second-order time derivative of a mode with a first-order time derivative. In the non-relativistic limit,
the second-order time derivative of Eq.~(\ref{Klein_Gordon_curved}) acting on $G(\theta)$ reads
\begin{equation}\label{Klein_Gordon_curved_approximation_positive_negative_frequencies}
-  \partial_0^2 G(\theta)= \frac{mc^2}{\hbar}   \left[ 2 i \partial_0 - \frac{mc^2}{\hbar} + \frac{mc^2}{\hbar}\mathcal{O} (\epsilon^2) \right] G(\theta) ,
\end{equation}
which is the equivalent of Eq.~(\ref{Klein_Gordon_approximation_positive_negative_frequencies}) in curved spacetime. By using Eq.~(\ref{Klein_Gordon_curved_approximation_positive_negative_frequencies}) in Eq.~(\ref{Klein_Gordon_curved}) for $G(\theta)$, we obtain
\begin{equation} \label{Schrodinger_curved_error}
i \hbar \partial_0 G(\theta) = [ H_\text{S} + mc^2 \mathcal{O} (\epsilon^2)] G(\theta),
\end{equation}
which leads to the Schr\"{o}dinger equation (\ref{Schrodinger_curved}). The error associated to such approximation reads
\begin{equation} \label{H_S_h_CKG_error}
H_\text{S} - h_\text{CKG} \sim \epsilon^2 mc^2.
\end{equation}

Equation (\ref{cKG_scalar_curved_product_G_nonrelativistic}) can be proven from Eqs.~(\ref{scalar_curved_modes_tilde}), (\ref{KG_curved_scalar_product_2}) and by replacing the frequencies with $mc^2/\hbar$. The relative error associated to such approximation is of the order $\epsilon$ as in Eq.~(\ref{scalar_product_error_order})
\begin{align} \label{scalar_product_error_order_curved}
\frac{( G(\theta), G(\theta') )_\text{CKG}}{\frac{2 m}{\hbar^2} ( G(\theta), G(\theta') )_{L^2_S(\mathbb{R}^3)}} - 1 \sim \epsilon.
\end{align}

Finally, the interacting theory can be described similarly to Sec.~\ref{Minkowski_spacetime_Scalar_field}. The only modification from the free theory is given by wave functions $\tilde{\Phi}_n (\bm{\theta}_n, T)$ that are now time dependent and, hence, generate an extra term in the Schr\"{o}dinger equation
\begin{align}\label{Schrodinger_n_particles_curved}
& i \hbar \partial_0  \Phi_n (T, \textbf{X}_n)   \approx \sum_{l=1}^n \left\lbrace \frac{m c^2}{2} \left[ 1 - \frac{g_{00}(\vec{X}_l)}{c^2} \right] \right. \nonumber \\
& \left. + \frac{\hbar^2 g_{00}(\vec{X}_l)}{2 m c^2} \nabla^2_{\vec{X}_l} \right\rbrace \Phi_n (T, \textbf{X}_n)+ \left( \frac{2 m}{\hbar^2} \right)^{n/2}  \nonumber \\
& \times  \sum_{\bm{\theta}_n}  \sum_{m=0}^\infty   \sum_{\bm{\theta}'_m} \langle \bm{\theta}_n | \hat{V}(T) | \bm{\theta}'_m \rangle \tilde{\Phi}_m (\bm{\theta}'_m, T)  \prod_{l=1}^n G(\theta_l,T,\vec{X}_l) .
\end{align}
In order to obtain such result, we consider the following condition
\begin{equation}\label{non_relativistic_potential_curved}
\langle \bm{\theta}_n | \hat{V}(T) | \bm{\theta}'_m \rangle  \sim \epsilon m c^2.
\end{equation}
Then by following the same arguments of Sec.~\ref{Minkowski_spacetime_Scalar_field} we obtain Eq.~(\ref{Schrodinger_n_particles_curved}). 

\subsection{Dirac field}\label{NonMinkowksi_spacetime_Dirac_field}

Here, we work in the $(T,\vec{X})$ frame with a Dirac field $\hat{\Psi}$. We show how non-relativistic single-particles can be represented by the space of positive frequency modes and through an inner product that is metric dependent. This result is similar to Sec.~\ref{Minkowski_spacetime_Dirac_field}. However, the single-particle representation is no more equivalent to the familiar position representation in $\mathbb{C}^2 \otimes L^2(\mathbb{R}^3)$.

We also show that non-relativistic Fock states are approximately solutions of a Schr\"{o}dinger equation that is different from the one obtained in Sec.~\ref{NonMinkowksi_spacetime_Scalar_field} for scalar fields. Such difference  is noticeable at any order, unless the metric is almost flat and the limit $g_{\mu\nu} \rightarrow \eta_{\mu\nu}$ is controlled by the non-relativistic parameter $\epsilon$. In that case, the difference between the scalar and Dirac Hamiltonians is not vanishing only at some orders. We discuss the situation in which these orders differ from the one seen for the Minkowski case [Eq.~(\ref{h_KG_h_M_error})].

The present section is organized as the previous ones. We start from the free theory. We provide the fully relativistic theory of particles and then we consider the non-relativistic limit. Finally we introduce an interaction through a potential term in the Schr\"{o}dinger equation.

The free field $\hat{\Psi}$ is solution of the curved spacetime Dirac equation due to Fock and Weyl --- see for instance \cite{collas_klein_2019} --- which reads
\begin{equation} \label{Dirac_curved}
\left( i c e_\alpha{}^\mu \gamma^\alpha D_\mu  - \frac{m c^2}{\hbar} \right) \hat{\Psi} = 0
\end{equation}
with
\begin{align}
& D_\mu = \partial_\mu + \Gamma_\mu, & \Gamma_\mu (T,\vec{X}) = - \frac{1}{2} \sigma^{\alpha \beta} \omega_{\alpha \beta \mu},
\end{align}
the spin connection
\begin{equation} \label{spin_connection}
\omega_{\alpha \beta \mu} = \eta_{\alpha \gamma} e^\gamma{}_\nu (\partial_\mu e_\beta{}^\nu + \Gamma^\nu{}_{\mu\rho} e_\beta{}^\rho),
\end{equation}
the Christoffel symbols
\begin{equation}
\Gamma^\rho{}_{\mu\nu} = \frac{1}{2} g^{\rho\sigma} (\partial_\nu g_{\sigma\mu} + \partial_\mu g_{\nu\sigma} -\partial_\sigma g_{\mu\nu})
\end{equation}
and the generators of the Clifford algebra
\begin{equation}
\sigma^{\mu\nu} = \frac{1}{4} [\gamma^\mu, \gamma^\nu].
\end{equation}

The following product \cite{PhysRevD.22.1922} can be defined for any couple of solutions of Eq.~(\ref{Dirac_curved})
\begin{align}\label{Dirac_curved_scalar_product}
( \Psi, \Psi' )_{\mathbb{C}^4 \otimes L^2_D(\mathbb{R}^3)} = & c \int_{\mathbb{R}^3} d^3X \sqrt{-g(T,\vec{X})} e_\alpha{}^0 (T,\vec{X}) \nonumber\\
& \times \Psi^\dagger(T,\vec{X}) \gamma^0 \gamma^\alpha \Psi'(T,\vec{X}).
\end{align}
$( \Psi, \Psi' )_{\mathbb{C}^4 \otimes L^2_D(\mathbb{R}^3)}$ can be seen as the inner product of $\mathbb{C}^4 \otimes L^2(\mathbb{R}^3)$ but with a metric dependent measure. It has been proven \cite{PhysRevD.22.1922} that when the metric is static, $( \Psi, \Psi' )_{\mathbb{C}^4 \otimes L^2_D(\mathbb{R}^3)}$ is time independent for solutions of Eq.~(\ref{Dirac_curved}). Therefore, we consider the case in which condition (\ref{metric_constraint_static}) holds. 

As a consequence of condition (\ref{metric_constraint_static}), Eq.~(\ref{vierbein_constraint_static}) holds, together with
\begin{align}
& \partial_0 \Gamma^\rho{}_{\mu\nu} = 0, & \Gamma^0{}_{i j} = \Gamma^i{}_{0 j} = \Gamma^i{}_{j 0} = 0.
\end{align}
Correspondingly,
\begin{align}
& \partial_0 \omega_{\alpha\beta\mu} = 0, & \omega_{i j 0} = \omega_{i 0 j} = \omega_{0 i j} = 0,
\end{align}
which leads to
\begin{align} \label{omega_condition_Dirac}
& \partial_0 \Gamma_\mu = 0, & & \Gamma_0 = -\frac{1}{4} \omega_{0 i 0} \sigma^{0 i}, & & \Gamma_i = -\frac{1}{8} \omega_{j k i} \sigma^{j k}.
\end{align}
By taking in count Eq.~(\ref{gamma_matrices_hermitianity}), we also find out that $\Gamma_0$ is hermitian while $\Gamma_i$ antihermitian
\begin{align} \label{omega_condition_Dirac_hermitianity}
&  \Gamma^\dagger_0 = \Gamma_0, & \Gamma^\dagger_i = - \Gamma_i.
\end{align}
Moreover, Eq.~(\ref{Dirac_curved_scalar_product}) now reads
\begin{align}\label{Dirac_curved_scalar_product_2}
( \Psi, \Psi' )_{\mathbb{C}^4 \otimes L^2_D(\mathbb{R}^3)} = & \frac{1}{c} \int_{\mathbb{R}^3} d^3X \sqrt{-g(\vec{X})} e_0{}^0 (\vec{X}) \nonumber \\
& \times \Psi^\dagger(T,\vec{X}) \Psi'(T,\vec{X}),
\end{align}
thanks to Eqs.~(\ref{gamma_matrices_anticommutating_rule}) and (\ref{vierbein_constraint_static}).

In a static spacetime, the Hamiltonian associated to the curved Dirac equation (\ref{Dirac_curved}) is hermitian with respect to the scalar product $( \Psi, \Psi' )_{\mathbb{C}^4 \otimes L^2_D(\mathbb{R}^3)}$. Such Hamiltonian is defined from the curved Dirac equation (\ref{Dirac_curved}) for static spacetimes
\begin{equation}\label{Dirac_curved_static}
\left[ i c e_0{}^0 \gamma^0 (\partial_0 + \Gamma_0) + i c e_i{}^j \gamma^i (\partial_j + \Gamma_j)  - \frac{m c^2}{\hbar} \right] \hat{\Psi} = 0
\end{equation}
and reads
\begin{equation}
h_\text{NM} =  - i \hbar c^2 e^0{}_0 e_i{}^j \gamma^0 \gamma^i (\partial_j + \Gamma_j)+ mc^3 e^0{}_0 \gamma^0 - i \hbar \Gamma_0.
\end{equation}
Indeed, by acting with $ \hbar c e^0{}_0 \gamma^0$ on the left of Eq.~(\ref{Dirac_curved_static}) and using Eq.~(\ref{gamma_matrices_anticommutating_rule}), one obtains
\begin{equation}\label{Dirac_curved_Schrodinger}
i \hbar \partial_0 \hat{\Psi} = h_\text{NM} \hat{\Psi}. 
\end{equation}

The proof for the hermicity of $h_\text{NM}$ with respect to $( \Psi, \Psi' )_{\mathbb{C}^4 \otimes L^2_D(\mathbb{R}^3)}$ arises from the fact that $( \Psi, \Psi' )_{\mathbb{C}^4 \otimes L^2_D(\mathbb{R}^3)}$ is time independent for solutions of the curved Dirac equation (\ref{Dirac_curved}) and, hence, for solutions of Eq.~(\ref{Dirac_curved_Schrodinger}):
\begin{align}
0 = & i \hbar \frac{d}{dt} ( \Psi, \Psi' )_{\mathbb{C}^4 \otimes L^2_D(\mathbb{R}^3)} \nonumber \\
 = &  -( i \hbar \partial_0 \Psi, \Psi' )_{\mathbb{C}^4 \otimes L^2_D(\mathbb{R}^3)} + (\Psi,  i \hbar \partial_0 \Psi' )_{\mathbb{C}^4 \otimes L^2_D(\mathbb{R}^3)} \nonumber \\
 = & -( h_\text{NM} \Psi, \Psi' )_{\mathbb{C}^4 \otimes L^2_D(\mathbb{R}^3)} + (\Psi,  h_\text{NM} \Psi' )_{\mathbb{C}^4 \otimes L^2_D(\mathbb{R}^3)}.
\end{align}

The hermicity of $h_\text{NM}$ guarantees the separation of the field into positive and negative frequency modes
\begin{equation}\label{Psi}
 \hat{\Psi}(T,\vec{X}) =  \sum_\theta \left[ U(\theta,T,\vec{X}) \hat{C}(\theta) + V(\theta,T,\vec{X}) \hat{D}^\dagger(\theta) \right],
\end{equation}
with
\begin{subequations}\label{Dirac_curved_modes_tilde_positive_negative}
\begin{align} 
& U(\theta,T,\vec{X}) = e^{-i \Omega(\theta) T} \tilde{U}(\theta,\vec{X}), \\
 & V(\theta,T,\vec{X}) = e^{i \Omega(\theta) T} \tilde{V}(\theta,\vec{X}).
\end{align}
\end{subequations}

The single-particle space is generated by the $U(\theta)$ modes and is supplemented by the $( \Psi, \Psi' )_{\mathbb{C}^4 \otimes L^2_D(\mathbb{R}^3)}$ product. It can be noticed that even in the non-relativistic limit (\ref{non_relativistic_limit_curved}), such representation is not equivalent to $\mathbb{C}^2 \otimes L^2(\mathbb{R}^3)$, at variance with the flat case. This occurs for two reasons: $( \Psi, \Psi' )_{\mathbb{C}^4 \otimes L^2_D(\mathbb{R}^3)}$ is metric dependent and the curved Dirac equation (\ref{Dirac_curved}) in the non-relativistic limit (\ref{non_relativistic_limit_curved}) does not lead to vanishing spinorial components for $U(\theta)$ modes. The familiar NRQT prescription of position representation through the $\mathbb{C}^2 \otimes L^2(\mathbb{R}^3)$ space cannot be restored in the curved case ($g_{\mu\nu} \neq \eta_{\mu\nu}$).

Single particles are also described by the Hamiltonian $h_\text{NM}$. In this section, we want to find an approximation for $h_\text{NM}$ in the non-relativistic limit by following the same steps of Sec.~\ref{Minkowski_spacetime_Dirac_field}. For this reason, we are interested in a Klein-Gordon-like equation for $U(\theta)$. Such equation exists and reads \cite{Pollock:2010zz}
\begin{equation} \label{Klein_Gordon_curved_Dirac}
\left[ \frac{c^2}{\sqrt{-g}} D_\mu \left( \sqrt{-g} g^{\mu \nu} D_\nu   \right) - \left( \frac{mc^2}{\hbar}\right)^2 - \frac{c^2}{4} R \right] U(\theta) = 0,
\end{equation}
with $R$ as the Ricci scalar. In Appendix \ref{appendix}, we give a detailed proof for such identity. In the static case, Eq.~(\ref{Klein_Gordon_curved_Dirac}) reads
\begin{align}\label{Klein_Gordon_curved_Dirac_static}
& \left\lbrace c^2  g^{0 0} \left( \partial_0 + \Gamma_0 \right)^2  +  \frac{c^2}{\sqrt{-g}} \left(\partial_i + \Gamma_i \right) \left[ \sqrt{-g} g^{i j} \left( \partial_j + \Gamma_j \right)  \right] \right. \nonumber \\
& \left. - \left( \frac{mc^2}{\hbar}\right)^2 - \frac{c^2}{4} R \right\rbrace U(\theta) = 0.
\end{align}
By using the curved Dirac equation (\ref{Dirac_curved_Schrodinger}) for $U(\theta)$ on Eq.~(\ref{Klein_Gordon_curved_Dirac_static}), we obtain
\begin{align}\label{Klein_Gordon_curved_Dirac_static_2}
& -\hbar^2 \partial_0^2 U(\theta) = \left\lbrace  \frac{\hbar^2 g_{0 0} }{\sqrt{-g}} \left(\partial_i + \Gamma_i \right) \left[ \sqrt{-g} g^{i j} \left( \partial_j + \Gamma_j \right)  \right] \right. \nonumber \\
& \left. - \frac{g_{0 0}}{c^2} \left[(mc^2)^2 + \frac{(\hbar c)^2}{4} R \right] - i 2 \hbar  \Gamma_0 h_\text{NM} + \hbar^2 \Gamma^2_0 \right\rbrace U(\theta).
\end{align}
In this way, we have been able to find $h_\text{NM}$ squared. Indeed, by using Eq.~(\ref{Dirac_curved_Schrodinger})  for $U(\theta)$ and Eq.~(\ref{Klein_Gordon_curved_Dirac_static_2}), we obtain
\begin{equation} \label{Klein_Gordon_curved_Dirac_static_3}
-\hbar^2 \partial_0^2 U(\theta) =  h_\text{NM}^2 U(\theta),
\end{equation}
and
\begin{align}\label{h_NM_2}
& h_\text{NM}^2 =   \frac{\hbar^2 g_{0 0} }{\sqrt{-g}} \left(\partial_i + \Gamma_i \right) \left[ \sqrt{-g} g^{i j} \left( \partial_j + \Gamma_j \right)  \right] \nonumber \\
&  - \frac{g_{0 0}}{c^2} \left[ (mc^2)^2 + \frac{(\hbar c)^2}{4} R \right] - i 2 \hbar  \Gamma_0 h_\text{NM} + \hbar^2 \Gamma^2_0 .
\end{align}

The second-order time derivative of Eq.~(\ref{Klein_Gordon_curved_Dirac}) is the same of Eq.~(\ref{Klein_Gordon_curved}). Moreover, Eq.~(\ref{Klein_Gordon_curved_approximation_positive_negative_frequencies}) is valid also for Dirac modes $U(\theta)$ in the non-relativistic limit. For these reasons, Eq.~(\ref{Klein_Gordon_curved_Dirac_static_3}) reads
\begin{equation} \label{Klein_Gordon_curved_Dirac_static_3_approximation}
mc^2 \left[ 2 i \hbar \partial_0 - mc^2 + mc^2 \mathcal{O} (\epsilon^2)  \right] U(\theta) = h_\text{NM}^2 U(\theta).
\end{equation}
If we now define the Hamiltonian
\begin{equation}\label{H_cD_h_nM}
H_\text{D} = \frac{h_\text{NM}^2}{2 mc^2}  + \frac{mc^2}{2},
\end{equation}
Eq.~(\ref{Klein_Gordon_curved_Dirac_static_3_approximation}) reads
\begin{equation}\label{Dirac_curved_Schrodinger_nonrelativistic_error}
i \hbar \partial_0 U(\theta) = [ H_\text{D} + mc^2 \mathcal{O} (\epsilon^2)] U(\theta).
\end{equation}
Equation (\ref{Dirac_curved_Schrodinger_nonrelativistic_error}) leads to the Schr\"{o}dinger equation
\begin{equation}\label{Dirac_curved_Schrodinger_nonrelativistic}
i \hbar \partial_0 U(\theta) \approx H_\text{D} U(\theta),
\end{equation}
with an error given by
\begin{equation} \label{H_D_h_nM_error}
H_\text{D} - h_\text{NM} \sim \epsilon^2 mc^2.
\end{equation}
From Eq.~(\ref{H_cD_h_nM}) one can see that the Hamiltonian $H_\text{D}$ is hermitian with respect to $( \Psi, \Psi' )_{\mathbb{C}^4 \otimes L^2_D(\mathbb{R}^3)}$ and can be used for the time evolution of non-relativistic states.

\begin{table}
\begin{tabular}{c|c|c|}
 & QFTCS & NRQTCS \\
\hline
$\langle \Psi | \Psi' \rangle$ & $(\Psi_1, \Psi'_1)_{\mathbb{C}^4 \otimes L^2_S(\mathbb{R}^3)}$ & $(\Psi_1, \Psi'_1)_{\mathbb{C}^4 \otimes L^2_S(\mathbb{R}^3)}$\\
\hline
Hamiltonian & $h_\text{NM}$ & $H_\text{D}$ \\
\hline
\end{tabular}
\caption{Inner product (first line) and Hamiltonian (second line) for free Dirac single-particles in curved spacetime. The left column is for the fully relativistic theory (QFTCS), while the right one is for the non-relativistic limit (NRQTCS).} \label{table_Dirac_nonMinkowski}
\end{table}

In summary, single-particles are described by the inner product $( \Psi, \Psi' )_{\mathbb{C}^4 \otimes L^2_D(\mathbb{R}^3)}$. The time evolution of single-particles is given by the Hamiltonian $h_\text{NM}$, which, in the non-relativistic limit, can be replaced by $H_\text{D}$. These results are shown schematically by Table \ref{table_Dirac_nonMinkowski}.

By comparing Eq.~(\ref{h_NM_2}) with Eq.~(\ref{Schrodinger_equation_tilde_Hamiltonian}) we can write
\begin{equation}
h_\text{NM}^2 = H_\text{CKG} + 2 m c^2 \Delta H
\end{equation}
and, hence
\begin{equation}\label{H_D_H_S_Delta_H}
H_\text{D} = H_\text{S} + \Delta H,
\end{equation}
with
\begin{align}
\Delta H = & \frac{\hbar^2 g_{0 0}}{2 mc^2} \left\lbrace \frac{[ \partial_i ( \sqrt{-g} g^{i j} \Gamma_j )]}{\sqrt{-g}} +  g^{i j} \Gamma_i ( 2 \partial_j + \Gamma_j ) - \frac{R}{4} \right\rbrace \nonumber \\
& - i \frac{\hbar}{mc^2} \Gamma_0 h_\text{NM} + \frac{\hbar^2}{2 mc^2} \Gamma^2_0.
\end{align}

For a non-flat metric ($g_{\mu \nu} \neq \eta_{\mu \nu}$), the difference between $H_\text{S}$ and $H_\text{D}$ is non-vanishing. At variance with the flat case, the spinorial decoupling does not occur and Dirac particles evolve differently from scalar states.

For Minkowski spacetimes ($g_{\mu \nu} = \eta_{\mu \nu}$), $\Delta H $ is identically vanishing and the difference between scalar and Dirac field is detectable only at order $\epsilon^2$ [Eq.~(\ref{h_KG_h_M_error})]. We wonder if this is also true for a quasi-flat spacetime ($g_{\mu \nu} \approx \eta_{\mu \nu}$). By considering the limit $g_{\mu \nu} \rightarrow \eta_{\mu \nu}$ regulated through the non-relativistic parameter $\epsilon$, different scenarios occur for different orders of magnitude of $\Delta H / (mc^2)$ with respect to $\epsilon$. For instance, if $\Delta H$ is of order lower than $\epsilon^2 m c^2$, the difference between $h_\text{CKG}$ and $h_\text{NM}$ is also of order lower than $\epsilon^2 m c^2$. In that case, one can distinguish between scalar and Dirac fields with less precision than the one needed for the flat case [Eq. (\ref{h_KG_h_M_error})]. Additionally, if $\Delta H$ is of order $\epsilon m c^2$, its contribution in Eq.~(\ref{H_D_H_S_Delta_H}) has the same magnitude of any leading correction to the mass energy $m c^2$. In other words, scalar and Dirac particles evolve differently already in the first non-trivial order of their Schr\"{o}dinger equation.

For completeness we provide the non-relativistic theory for states different from the single-particles. A general Fock state $| \Psi \rangle$ is represented in the Schr\"{o}dinger picture by
\begin{equation}\label{Dirac_curved_wavefunction_general}
\Psi^{\bm{\alpha}_n}_n (T, \textbf{X}_n)   =  \sum_{\bm{\theta}_n} \tilde{\Psi}_n (\bm{\theta}_n) \prod_{l=1}^n U^{\alpha_l}(\theta_l, T, \vec{X}_l).
\end{equation}
where $ \tilde{\Psi}_n (\bm{\theta}_n)$ comes from the decomposition of $| \Psi \rangle$ in the Fock space, as in Eq.~(\ref{general_Fock_Dirac_expansion})
\begin{align} \label{general_Fock_Dirac_curved_expansion}
| \Psi \rangle  = & \sum_{n=0}^\infty \sum_{\bm{\theta}_n} \tilde{\Psi}_n (\bm{\theta}_n) | \bm{\theta}_n \rangle.
\end{align}
The inner product between two states $| \Psi \rangle$, $| \Psi' \rangle$ can be achieved through a generalization of $( \Psi, \Psi' )_{\mathbb{C}^4 \otimes L^2_D(\mathbb{R}^3)}$ for states with indefinite number of particles
\begin{equation}
\langle \Psi | \Psi' \rangle = \sum_{n=0}^\infty (\Psi_n, \Psi'_n)_{\mathbb{C}^{4n} \otimes L^2_D(\mathbb{R}^{3n})},
\end{equation}
with
\begin{subequations}
\begin{align}
& (\Psi_n, \Psi'_n)_{\mathbb{C}^{4n} \otimes L^2_D(\mathbb{R}^{3n})} = \frac{1}{c^n} \sum_{\bm{\alpha}_n} \int_{\mathbb{R}^{3n}} d^{3n} \textbf{X}_n  \nonumber \\
& \times \left[ \prod_{l=1}^n \sqrt{-g(\vec{X}_l)} e_0{}^0 (\vec{X}_l) \right] [\Psi^*_n (T, \textbf{X}_n)]^{\bm{\alpha}_n} [\Psi'_n (T, \textbf{X}_n)]^{\bm{\alpha}_n},
\end{align}
\begin{equation}
(\Psi_0, \Psi'_0)_{\mathbb{C}^0 \otimes L^2_D(\mathbb{R}^0)} = \Psi_0^* \Psi'_0.
\end{equation}
\end{subequations}

The Schr\"{o}dinger equation for $\Psi_n (T, \textbf{X}_n)$ is equivalent to Eq.~(\ref{Schrodinger_free_mass_n_particles_curved}) with extra terms coming from a non-vanishing $\Delta H$:
\begin{align}\label{Schrodinger_free_mass_n_particles_Dirac_curved}
& i \hbar \partial_0  \Psi^{\bm{\alpha}_n}_n (T, \textbf{X}_n) \approx \sum_{l=1}^n \left\lbrace \frac{m c^2}{2} \left[ 1 - \frac{g_{00}(\vec{X}_l)}{c^2} \right] \right. \nonumber \\
& \left. + \frac{\hbar^2 g_{00}(\vec{X}_l)}{2 m c^2} \nabla^2_{\vec{X}_l} \right\rbrace \Psi^{\bm{\alpha}_n}_n (T, \textbf{X}_n)  \nonumber \\
& + \sum_{l=1}^n \Delta H(\vec{X}_l)^{\alpha_l}{}_{\beta_l} \Psi^{\alpha_1 \dots \beta_l \dots \alpha_n}_n (T, \textbf{X}_n).
\end{align}

Finally, regarding the theory with interaction, we may use the same arguments of Sec.~\ref{Minkowski_spacetime_Dirac_field} to conclude that the resulting modification is an extra term in the Schr\"{o}dinger equation (\ref{Schrodinger_free_mass_n_particles_Dirac_curved}):
\begin{align}\label{Schrodinger_free_mass_n_particles_Dirac_curved_interaction}
& i \hbar \partial_0  \Psi^{\bm{\alpha}_n}_n (T, \textbf{X}_n)  \approx \sum_{l=1}^n \left\lbrace \frac{m c^2}{2} \left[ 1 - \frac{g_{00}(\vec{X}_l)}{c^2} \right] \right. \nonumber \\
& \left. + \frac{\hbar^2 g_{00}(\vec{X}_l)}{2 m c^2} \nabla^2_{\vec{X}_l} \right\rbrace \Psi^{\bm{\alpha}_n}_n (T, \textbf{X}_n) + \sum_{l=1}^n \Delta H(\vec{X}_l)^{\alpha_l}{}_{\beta_l} \nonumber \\
& \times \Psi^{\alpha_1 \dots \beta_l \dots \alpha_n}_n (T, \textbf{X}_n) + \sum_{\bm{\theta}_n} \sum_{m=0}^\infty   \sum_{\bm{\theta}'_m} \langle \bm{\theta}_n | \hat{V}(T) | \bm{\theta}'_m \rangle \nonumber \\
 & \times  \tilde{\Psi}_m (\bm{\theta}'_m, T) \prod_{l=1}^n U^{\alpha_l}(\theta_l, T, \vec{X}_l) .
\end{align}

\section{Rindler frame}\label{Rindler_frame}

As an example of noninertial frame $(T,\vec{X})$, here we consider a  Rindler frame, such that
\begin{equation} \label{Rindler_metric}
g_{\mu\nu}(T,\vec{X}) = \text{diag} \left( -c^2 e^{2aX}, e^{2aX}, 1, 1 \right),
\end{equation}
where $\alpha =c^2 a$ is the acceleration along the $X$ axis. Without loss of generality we consider $\alpha>0$. We adopt the theory of Sec.~\ref{NonMinkowksi_spacetime} to derive the non-relativistic limit for particle states in the Rindler spacetime. We discuss the cases in which the time evolution of scalar and Dirac field differs.

\subsection{Scalar field}\label{Rindler_frame_Scalar_field}

Here we work with the scalar field $\hat{\Phi}(T,X)$. Firstly we derive the non-relativistic limit of single particles. We compute the Schr\"{o}dinger equation and the inner product of non-relativistic single-particles. Thanks to the particular form of the Rindler metric, we conclude that such product can be approximated by the $L^2(\mathbb{R}^3)$ inner product. In this way, we show that non-relativistic Rindler particles can be equivalently treated as if they were in a flat spacetime, but with a modified free Schr\"{o}dinger equation. Such modifications depend on the magnitude of the acceleration. By considering an $\alpha$ that is constrained by the non-relativistic limit, we show how the Schr\"{o}dinger equation is further approximated by the familiar Schr\"{o}dinger-Newton equation. 

In the case of scalar fields in Rindler spacetime, Eq.~(\ref{Klein_Gordon_curved}) reads
\begin{equation}\label{Rindler_Klein_Gordon}
\left\lbrace - \partial_0^2 + c^2 \partial_1^2 +   c^2 e^{2 a X} \left[ \partial_2^2 + \partial_3^2 - \left( \frac{mc}{\hbar} \right)^2 \right] \right\rbrace \hat{\Phi} = 0.
\end{equation}
An explicit decomposition of $\hat{\Phi}$ is known \cite{RevModPhys.80.787} and reads
\begin{subequations}
\begin{align}
\hat{\Phi}(T,X) = & \int_0^{+\infty} d\Omega \int_{\mathbb{R}^2} d^2 k_\perp \left[ F(\Omega,\vec{k}_\perp,T,\vec{X})\hat{A}(\Omega,\vec{k}_\perp) \right. \nonumber \\
& \left. +  F^*(\Omega,\vec{k}_\perp,T,\vec{X})\hat{B}^\dagger(\Omega,\vec{k}_\perp) \right], \label{Rindler_scalar_decomposition}
\end{align}
\begin{align}
F(\Omega,\vec{k}_\perp,T,\vec{X}) = & \tilde{F}(\Omega,\vec{k}_\perp,X) e^{ i \vec{k}_\perp \cdot \vec{X}_\perp - i \Omega T }, \label{F_Rindler} 
\end{align}
\begin{align}
& \tilde{F}(\Omega,\vec{k}_\perp,X) = \frac{1}{2 \pi^2} \sqrt{ \frac{\hbar}{a} \sinh \left( \frac{\pi \Omega}{c a} \right) } \nonumber \\
 & \times K_{i \Omega / (c a)} \left( \sqrt{c^2 k_\perp^2 + \left(\frac{mc^2}{\hbar}\right)^2} \frac{e^{aX}}{c a} \right), \label{F_tilde_Rindler}
\end{align}
\end{subequations}
with $\vec{X}_\perp = (Y, Z)$ and where $K_\zeta (\xi)$ is the modified Bessel function of the second kind. It can be noticed that in the Rindler spacetime
\begin{equation}
g_{00} = c \sqrt{-g}
\end{equation}
and, hence,
\begin{align}\label{Rindler_Klein_Gordon_product}
& ( \Phi, \Phi' )_\text{CKG} = ( \Phi, \Phi' )_\text{KG}, & ( \Phi, \Phi' )_{L^2_S(\mathbb{R}^3)} = ( \Phi, \Phi' )_{L^2(\mathbb{R}^3)}.
\end{align}
The $F(\Omega,\vec{k}_\perp)$ modes defined in Eq.~(\ref{F_Rindler}) are orthonormal with respect to the $( \Phi, \Phi' )_\text{KG}$ scalar product and, hence, orthonormal with respect to $( \Phi, \Phi' )_\text{CKG}$.

For $\Omega$ such that Eq.~(\ref{non_relativistic_limit_curved}) holds, $F(\Omega,\vec{k}_\perp)$ is approximately solution of
\begin{equation} \label{Schrodinger_Rindler_scalar_F}
i \hbar \partial_0 F(\Omega,\vec{k}_\perp) \approx H_\text{S} F(\Omega,\vec{k}_\perp),
\end{equation}
with, in this case,
\begin{equation}\label{H_nM_Rindler}
H_\text{S} = - \frac{\hbar^2}{ 2 m} \left[ \partial_1^2  + e^{2aX} \left( \partial_2^2 + \partial_3^2 \right) \right] + \frac{mc^2}{2} \left(  1  + e^{2aX}  \right).
\end{equation}
Equations (\ref{Schrodinger_Rindler_scalar_F}) and (\ref{H_nM_Rindler}) can be checked by using Eq.~(\ref{Rindler_metric}) in Eqs.~(\ref{Schrodinger_curved}) and (\ref{H_nM}).

Moreover, in the non-relativistic limit, the Klein-Gordon product can be approximated by Eq.~(\ref{cKG_scalar_curved_product_G_nonrelativistic}). In the case of Rindler modes,
\begin{align}\label{cKG_scalar_curved_product_F_nonrelativistic}
& ( F(\Omega,\vec{k}_\perp), F(\Omega',\vec{k}'_\perp) )_\text{CKG} \nonumber \\ \approx  &\frac{2 m}{\hbar^2} ( F(\Omega,\vec{k}_\perp), F(\Omega',\vec{k}'_\perp) )_{L^2_S(\mathbb{R}^3)}.
\end{align}
Thanks to Eq.~(\ref{Rindler_Klein_Gordon_product}), Eq.~(\ref{cKG_scalar_curved_product_F_nonrelativistic}) can be also replaced by
\begin{align}\label{cKG_scalar_curved_product_F_nonrelativistic_2}
& ( F(\Omega,\vec{k}_\perp), F(\Omega',\vec{k}'_\perp) )_\text{CKG} \nonumber \\
 \approx & \frac{2 m}{\hbar^2} ( F(\Omega,\vec{k}_\perp), F(\Omega',\vec{k}'_\perp) )_{L^2(\mathbb{R}^3)}.
\end{align}
This means that non-relativistic Rindler single-particles can be treated identically to Minkowski particles, but with different free Schr\"{o}dinger equation (\ref{Schrodinger_Rindler_scalar_F}).

Such result is independent of the orders of magnitude for $\vec{X}$, $\vec{k}_\perp$ and $a$. Therefore, Eq.~(\ref{Schrodinger_Rindler_scalar_F}) can be considered in all physical scenarios where the energy of the system is non-relativistic, while $\vec{X}$, $\vec{k}_\perp$ and $a$ can assume any values. 

It can be proven that a further approximation for Eq.~(\ref{Schrodinger_Rindler_scalar_F}) holds if, together with the non-relativistic limit (\ref{non_relativistic_limit_curved}), one considers the following orders of magnitude for the variables $X$ and $\vec{k}_\perp$ and the parameter $a$
\begin{align}\label{local_limit}
& a |X| \sim \epsilon , & \frac{ \hbar |\vec{k}_\perp|}{m c} \sim \epsilon^{1/2},&  & \frac{\hbar a}{m c} \sim \epsilon^{3/2} ,
\end{align}
where $\epsilon $ is defined in Eq.~(\ref{non_relativistic_order}) and represents the ratio between the non-relativistic energy $E=\hbar \Omega - mc^2$ and the mass energy $mc^2$. The limits expressed by Eqs.~(\ref{non_relativistic_limit_curved}) and (\ref{local_limit}) can alternatively be obtained from $c \rightarrow \infty$, with $X$, $\vec{k}_\perp$, $\alpha$, $E$ fixed.

The condition $|X|  \ll 1/a $ means that we consider states with wave functions that are mostly localized in a region of spacetime that is close to the accelerated observer position $X=0$ with respect to the Rindler length scale $1/a$. In other words, the limit $a |X| \ll 1 $ can be identified with a locality condition such that curvature effects are considered small. Indeed, it is straightforward to see that when $a|X| \ll 1$, $g_{\mu\nu}$ is almost flat. For this reason we name Eq.~(\ref{local_limit}) \textit{quasi-inertial limit}. The fact that $a |X|$ goes to zero with the same order of $\epsilon$ means that
\begin{equation}
U_\text{g} = m  \alpha X
\end{equation}
has the same magnitude of $E$ (i.e., $U_\text{g} \sim E \sim \epsilon mc^2$) and can therefore be regarded as a non-relativistic energy. We anticipate that $U_\text{g}$ represents the potential energy for the approximated Schr\"{o}dinger equation in the limits (\ref{non_relativistic_limit_curved}) and (\ref{local_limit}). The condition $\hbar^2 k^2_\perp /  m^2 \sim \epsilon$ can also be interpreted as a non-relativistic condition for the transverse kinetic energy $\hbar^2 k^2_\perp / (2 m) \sim E$.

When Eqs.~(\ref{non_relativistic_limit_curved}) and (\ref{local_limit}) hold, Eq.~(\ref{F_tilde_Rindler}) can be approximated by
\begin{align}\label{F_tilde_Rindler_Hankel_2}
& \tilde{F}(\Omega,\vec{k}_\perp,X) \approx \frac{\hbar^{5/3}  }{2^{7/6} \pi  a^{1/6} (m c)^{1/3}} \text{Ai} \left( 2^{1/3} \left(\frac{m c}{\hbar a} \right)^{2/3} \right. \nonumber \\
& \left. \times \left[ \frac{\hbar^2 k_\perp^2}{2 m^2 c^2} + a X - \left( \frac{\hbar \Omega}{m c^2} - 1  \right) \right] \right),
\end{align}
where $\text{Ai}$ is the Airy function. The proof for Eq.~(\ref{F_tilde_Rindler_Hankel_2}) is provided by Appendix \ref{Appendix_2}. From Eq.~(\ref{F_tilde_Rindler_Hankel_2}), one can see that $F(\Omega,\vec{k}_\perp)$ is approximately solution of
\begin{equation}\label{Schrodinger_Rindler_scalar_local}
i \hbar \partial_0 F(\Omega,\vec{k}_\perp) \approx H_\text{QI} F(\Omega,\vec{k}_\perp),
\end{equation}
with
\begin{equation}\label{H_QI}
H_\text{QI} = - \frac{\hbar^2}{ 2 m} \left( \partial_1^2  + \partial_2^2 + \partial_3^2 \right) + mc^2  +  U_\text{g}.
\end{equation}
Indeed, by knowing that the Airy function is solution of the differential equation $\text{Ai}''(x) = x \text{Ai}(x)$, one can prove from Eq.~(\ref{F_tilde_Rindler_Hankel_2}) that
\begin{align}\label{Schrodinger_Rindler_local}
& \partial_1^2 F(\Omega,\vec{k}_\perp,T,\vec{X}) \approx 2 \left(\frac{m c}{\hbar} \right)^2 \left[ \frac{\hbar^2 k_\perp^2}{2 m^2 c^2} + a X \right. \nonumber \\
& \left. - \left( \frac{\hbar \Omega}{m c^2} - 1  \right) \right] F(\Omega,\vec{k}_\perp,T,\vec{X}) ,
\end{align}
which, together with Eq.~(\ref{F_Rindler}), proves Eq.~(\ref{Schrodinger_Rindler_scalar_local}).

Equation (\ref{Schrodinger_Rindler_scalar_local}) is a Schr\"{o}dinger-Newton equation with a mass term $ mc^2 $ and a potential energy $U_\text{g}$ generated by an uniform gravitational force $m \alpha$ along the $X$ axes. This can be interpreted as the fact that an accelerated frame is locally equivalent to an observer that experiences a gravitational force. The result is hence expected by the equivalence principle of general relativity and the limits that we have considered.

The error associated to Eq.~(\ref{Schrodinger_Rindler_scalar_local}) approximating Eq.~(\ref{Schrodinger_Rindler_scalar_F}) can be obtained by evaluating the difference between the two Hamiltonians $H_\text{S}$, $H_\text{QI}$ acting on $F(\Omega,\vec{k}_\perp)$:
\begin{equation}
H_\text{S} - H_\text{QI} = \frac{\hbar^2 k_\perp^2}{ 2 m} (e^{2aX} - 1)  + mc^2 \left( \frac{e^{2aX} - 1}{2} - aX \right).
\end{equation}
For non-relativistic modes $F(\Omega,\vec{k}_\perp)$ and in the quasi-inertial limit (\ref{local_limit}), $H_\text{S} - H_\text{QI}$ acts on $F(\Omega,\vec{k}_\perp)$ with the following leading order
\begin{equation}\label{H_S_H_QI}
H_\text{S} - H_\text{QI} \sim \epsilon^2 mc^2.
\end{equation}
By comparing Eq.~(\ref{H_S_H_QI}) with Eq.~(\ref{H_S_h_CKG_error}), one notices that the errors associated to Eq.~(\ref{Schrodinger_Rindler_scalar_local}) are of the same orders of Eq.~(\ref{Schrodinger_Rindler_scalar_F}). Therefore, no reason to prefer the Hamiltonian $H_\text{S}$ over $ H_\text{QI}$ exists: they can be considered equivalent in the non-relativistic quasi-inertial regime. Moreover, the difference between the Hamiltonian $H_\text{QI}$ and the exact fully-relativistic $h_\text{CKG}$ reads
\begin{equation} \label{H_QI_h_CKG_error}
H_\text{QI} - h_\text{CKG} \sim \epsilon^2 mc^2.
\end{equation}
Equation (\ref{H_QI_h_CKG_error}) gives an esteem of the GR corrections to the Schr\"{o}dinger-Newton equation (\ref{Schrodinger_Rindler_scalar_local}) for scalar fields.

It can be noticed that a similar result holds when one consider the following limit
\begin{align}\label{local_limit_alternative}
& a |X| \sim \epsilon , & \frac{ \hbar |\vec{k}_\perp|}{m c} \sim \epsilon^{1/2} ,&  & \frac{\hbar a}{m c} \sim \epsilon.
\end{align}
Equation (\ref{local_limit_alternative}) can be identified with the quasi-inertial limit (\ref{local_limit}) considered above, but with a different order of magnitude for $\hbar a / (m c)$ with respect to the non-relativistic limit. Moreover, Eq.~(\ref{local_limit_alternative}) cannot be obtained from the limit $c \rightarrow \infty$, with $X$, $\vec{k}_\perp$, $\alpha$, $E$ fixed. A larger acceleration is required here, as opposed to the limit (\ref{local_limit}). We, hence, name Eq.~(\ref{local_limit_alternative}) \textit{high acceleration limit}.

By following a proof similar to Appendix \ref{Appendix_2}, one can show that in the high acceleration limit (\ref{local_limit_alternative}), the Rindler modes $F(\Omega,\vec{k}_\perp)$ can be approximated by Eq.~(\ref{F_tilde_Rindler_Hankel_2}) and, hence, are approximately solutions of Eq.~(\ref{Schrodinger_Rindler_scalar_local}). The only difference with the previous case relies on the fact that the argument of the Airy function is of order $\epsilon^{1/3}$. Equation (\ref{H_QI_h_CKG_error}) also holds in the high acceleration limit (\ref{local_limit_alternative}) and provides the error associated to the Schr\"{o}dinger-Newton equation (\ref{Schrodinger_Rindler_scalar_local}).

\subsection{Dirac field}

Here we discuss the case of Dirac fields $\hat{\Psi}$ in Rindler spacetime. We first review the non-relativistic limit of single-particles. We derive a Schr\"{o}dinger equation that is different from the scalar case.

We then show that in the quasi-inertial limit (\ref{local_limit}) such equation can be approximated by the Schr\"{o}dinger-Newton equation (\ref{Schrodinger_Rindler_scalar_local}). GR corrections to such equation are $\epsilon^{-1/2}$ times larger than the ones obtained for the scalar field. This means that Dirac fields are better candidates for detecting GR corrections to the Schr\"{o}dinger-Newton theory. Moreover, we show that the difference between scalar and Dirac Hamiltonians is $\epsilon^{-1/2}$ times larger than what we found for the Minkowski case [Eq.~(\ref{h_KG_h_M_error})]. In other words, the Rindler metric is able to enhance the distinguishability between scalar and Dirac fields.

Finally, by considering the high acceleration limit (\ref{local_limit_alternative}), we show that, in such case, GR corrections to the Schr\"{o}dinger-Newton theory are of the same order of the Newtonian gravitational potential. Therefore, the dynamics of Dirac particles is different from the scalar case already at the first non-trivial order. 

We consider a Dirac field $\hat{\Psi}$ that is solution to the Dirac equation (\ref{Dirac_curved}) in Rindler spacetime (\ref{Rindler_metric}). The explicit form of such equation can be given by computing the vierbein field $e_\alpha{}^\mu$ and the matrices $\Gamma_\mu$. The only non-vanishing components of $e_\alpha{}^\mu$, $\partial_\mu e_\alpha{}^\nu$, $\partial_\rho g_{\mu\nu}$, $\Gamma^\rho{}_{\mu\nu}$, $\omega_{\alpha \beta \mu}$ and $\Gamma_\mu$ are the following
\begin{subequations}
\begin{align}
& e_0{}^0(T,\vec{X}) = e^{-aX}, & & e_1{}^1(T,\vec{X}) = e^{-aX},\\
 & e_2{}^2(T,\vec{X}) = 1, &  & e_3{}^3(T,\vec{X}) = 1, \\
& \partial_1 e_0{}^0(T,\vec{X}) = -a e^{-aX}, & &  \partial_1 e_1{}^1(T,\vec{X}) = -a e^{-aX},\\
 & \partial_1 g_{00}(T,\vec{X}) = - 2 c^2 a e^{2aX}, & &  \partial_1 g_{11}(T,\vec{X}) = 2 a e^{2aX}, \\
& \Gamma^1{}_{00}(T,\vec{X}) = c^2 a, & &  \Gamma^0{}_{10}(T,\vec{X}) = a, \\
 & \Gamma^0{}_{01}(T,\vec{X}) = a,  & &  \Gamma^1{}_{11}(T,\vec{X}) = a, \\
& \omega_{100}(T,\vec{X}) = c^2 a, & &  \omega_{010}(T,\vec{X}) = -c^2 a,\\
& \Gamma_0(T,\vec{X}) = \frac{c^2 a}{2}  \gamma^0 \gamma^1.
\end{align}
\end{subequations}
Equation (\ref{Dirac_curved}) now reads
\begin{align} \label{Dirac_Rindler}
& \left[ i c \gamma^0 \partial_0 + i \frac{c a}{2} \gamma^1 +  i c \gamma^1 \partial_1  \right. \nonumber \\
& \left. + e^{aX} \left( i c \gamma^2 \partial_2 + i c \gamma^3 \partial_3 - \frac{m c^2}{\hbar} \right) \right] \hat{\Psi} = 0,
\end{align}
while the scalar product (\ref{Dirac_curved_scalar_product_2}) reads
\begin{equation}\label{Dirac_Rindler_scalar_product}
( \Psi, \Psi' )_{\mathbb{C}^4 \otimes L^2_D(\mathbb{R}^3)} = \int_{\mathbb{R}^3} d^3X e^{aX} \Psi^\dagger(T,\vec{X}) \Psi'(T,\vec{X}).
\end{equation}

In the non-relativistic limit (\ref{non_relativistic_limit_curved}), $U(\theta)$ is approximately solution of the Schr\"{o}dinger equation (\ref{Dirac_curved_Schrodinger_nonrelativistic}) that reads:
\begin{equation}\label{Schrodinger_Rindler_Dirac}
i \hbar \partial_0 U(\theta) \approx ( H_\text{S} + \Delta H ) U(\theta),
\end{equation} 
where, in this case, $H_\text{S}$ is given by Eq.~(\ref{H_nM_Rindler}) and $\Delta H$ by
\begin{equation}\label{Schrodinger_Rindler_Dirac_Delta_H}
\Delta H =  - i \frac{\hbar a}{2 m} \gamma^0 \gamma^1 h_\text{NM} + \frac{(\hbar a)^2}{8 m},
\end{equation} 
\begin{align}\label{Schrodinger_Rindler_Dirac_h_NM}
& h_\text{NM} =  - \hbar c \gamma^0  \left[ i \frac{ca}{2} \gamma^1 +  i c \gamma^1 \partial_1  \right. \nonumber\\
& \left. + e^{aX} \left(i c \gamma^2 \partial_2 + i c \gamma^3 \partial_3 - \frac{mc^2}{\hbar} \right) \right] .
\end{align} 

It can be noticed that the Schr\"{o}dinger equation (\ref{Schrodinger_Rindler_Dirac}) differs from the scalar field case [Eq.~(\ref{Schrodinger_Rindler_scalar_F})]. The difference between the two Hamiltonians $H_\text{D}$ and $H_\text{S}$ is given by Eq.~(\ref{Schrodinger_Rindler_Dirac_Delta_H}), which, in the non-relativistic limit reads
\begin{equation}\label{Delta_H_Rindler_nonrelativistic}
\Delta H \approx  - i \frac{\hbar c^2 a}{2} \gamma^0 \gamma^1 + \frac{(\hbar a)^2}{8 m}.
\end{equation}
Equation (\ref{Delta_H_Rindler_nonrelativistic}) is generally non-vanishing. As already explained in Sec.~\ref{NonMinkowksi_spacetime_Dirac_field}, this occurs because the metric is not flat.

Different scenarios are possible when $a$ varies with respect to other dimensional quantities and $\epsilon$. For instance, in the case of quasi-inertial limit defined by Eq.~(\ref{local_limit}), the scalar product (\ref{Dirac_Rindler_scalar_product}) can be approximated by $\mathbb{C}^4 \otimes L^2(\mathbb{R}^3)$ inner product
\begin{equation}\label{Dirac_Rindler_scalar_product_approximation}
( \Psi, \Psi' )_{\mathbb{C}^4 \otimes L^2_D(\mathbb{R}^3)} \approx ( \Psi, \Psi' )_{\mathbb{C}^4 \otimes L^2(\mathbb{R}^3)}
\end{equation}
and the dynamics of the single-particles is reduced to the familiar Schr\"{o}dinger-Newton equation
\begin{equation}\label{Schrodinger_Newton_Rindler_Dirac}
i \hbar \partial_0 U(\theta) \approx H_\text{QI} U(\theta),
\end{equation} 
already defined for scalar particles by Eqs.~(\ref{Schrodinger_Rindler_scalar_local}) and (\ref{H_QI}).

Equation (\ref{Dirac_Rindler_scalar_product_approximation}) is due to the fact that in the quasi-inertial limit, wave functions are localized inside the region $a|X| \ll 1$, and, hence, $e^{aX} \approx 1$. Equation (\ref{Schrodinger_Newton_Rindler_Dirac}) can be proven by noticing that $\Delta H$, acting on non-relativistic states, is approximated by Eq.~(\ref{Delta_H_Rindler_nonrelativistic}) and, hence, in the quasi-inertial limit (\ref{local_limit}),
\begin{equation} \label{Delta_H_W_approximation_2}
\Delta H  \sim \epsilon^{3/2} mc^2,
\end{equation}
which is $\epsilon^{1/2}$ times smaller than the potential energy $U_\text{g} \sim  \epsilon mc^2$. This, together with the fact that in the quasi-inertial limit (\ref{local_limit}), $H_\text{S}$ can be replaced by $H_\text{QI}$ [Eq.~(\ref{H_S_H_QI})], leads to Eq.~(\ref{Schrodinger_Newton_Rindler_Dirac}).

In summary, Dirac modes are approximately solution of the same Schr\"{o}dinger equation for the scalar field and the scalar product is the same one defined for Dirac fields in Minkowski spacetime. Analogously, the scalar product for scalar fields is approximated by the $L^2(\mathbb{R}^3)$ inner product [Eq.~(\ref{cKG_scalar_curved_product_F_nonrelativistic_2})]. This means that non-relativistic quasi-inertial Dirac particles can be described identically to scalar states with the exception of spin degeneracy, as it occurs in the Minkowski spacetime.

From Eqs.~(\ref{H_D_h_nM_error}), (\ref{H_S_H_QI}) and (\ref{Delta_H_W_approximation_2}), one can notice that the errors associated to the Schr\"{o}dinger-Newton equation (\ref{Schrodinger_Newton_Rindler_Dirac}) are dominated by $\Delta H$ and read
\begin{equation} \label{H_QI_h_nM_error}
H_\text{QI} - h_\text{NM} \sim \epsilon^{3/2} mc^2.
\end{equation}
By comparing Eq.~(\ref{H_QI_h_nM_error}) with Eq.~(\ref{H_QI_h_CKG_error}), one can deduce that GR corrections to the Schr\"{o}dinger-Newton equation for Dirac fields are $\epsilon^{-1/2}$ times larger than the GR corrections for scalar fields. By increasing the experimental precision for energies up to the order of $\epsilon^{3/2} m c^2$, a the term proportional to $\gamma^0 \gamma^1$ [Eq.~(\ref{Delta_H_Rindler_nonrelativistic})] appears in the Dirac case, while nothing shows up for scalar fields.

By also comparing Eqs.~(\ref{H_QI_h_CKG_error}) and (\ref{H_QI_h_nM_error}), we find out that
\begin{equation} \label{h_CKG_h_nM_error}
h_\text{CKG} - h_\text{NM} \sim \epsilon^{3/2} mc^2,
\end{equation}
which means that the difference between scalar and Dirac Hamiltonians is visible at order $\epsilon^{3/2}$. Such order is lower than the one needed for the distinguishability between the two types of fields in the Minkowski spacetime [Eq.~(\ref{h_KG_h_M_error})]. The result is that in the Rindler frame, even in the quasi-inertial limit (\ref{local_limit}), it is easier to distinguish between scalar and Dirac fields than in the Minkowski spacetime.

A different scenario can be considered by changing the asymptotic behavior of $a$ with respect to the non-relativistic limit. For instance, by considering the high acceleration limit (\ref{local_limit_alternative}), we obtain
\begin{equation}\label{Delta_H_relativistic}
\Delta H \sim \epsilon mc^2.
\end{equation}
In this scenario, a non-vanishing $\Delta H$ appears at the same order of the Newtonian gravitational potential $\Delta H \sim U_\text{g}$. Therefore Eq.~(\ref{Schrodinger_Rindler_Dirac}) cannot be approximated by the Schr\"{o}dinger-Newton equation (\ref{Schrodinger_Newton_Rindler_Dirac}). Instead, we have
\begin{equation}
i \hbar \partial_0 U(\theta) \approx \left( H_\text{QI}  - i \frac{\hbar c^2 a}{2} \gamma^0 \gamma^1 \right) U(\theta).
\end{equation}

\begin{table}
\begin{tabular}{c|c|c|c||c|}
 & $a |X|$ & $\dfrac{\hbar |\vec{k}_\perp|}{m c} $ & $\dfrac{\hbar a}{m c}$ & $\dfrac{\Delta h}{mc^2}$ \\[1.5ex]
\hline
& & & & \\[-1em]
quasi-inertial limit (\ref{local_limit}) & $\epsilon$ & $\epsilon^{1/2}$ & $\epsilon^{3/2}$ & $\epsilon^{3/2}$ \\
\hline
& & & & \\[-1em]
high acceleration limit (\ref{local_limit_alternative}) & $\epsilon$ & $\epsilon^{1/2}$ & $\epsilon$ & $\epsilon$  \\
\hline
\end{tabular}
\caption{Asymptotic behavior with respect to the non-relativistic parameter $\epsilon$ for different limits. The quasi-inertial and the high acceleration limit are defined by, respectively, Eqs.~(\ref{local_limit}) and (\ref{local_limit_alternative}) in terms of the position $X$, the transverse momentum $\vec{k}_\perp$, the acceleration $\alpha = a c^2$ and $\epsilon$. The variable $\Delta h = h_\text{CKG} - h_\text{NM}$ is the difference between the scalar and Dirac Hamiltonians. The orders of $\Delta h $ for the two limits are shown in the last column. They are always lower than the order of $\Delta h $ in Minkowski spacetime $(h_\text{KG} - h_\text{M})/(mc^2) \sim \epsilon^2$ [Eq.~(\ref{h_KG_h_M_error})]. This means that lower precision is needed to distinguish between the time evolution of scalar and Dirac fields.} \label{table_Rindler_limits}
\end{table}

The difference between scalar and Dirac fields appears at the first-order correction to the mass energy term
\begin{equation}
h_\text{CKG} - h_\text{NM} \sim \epsilon mc^2,
\end{equation}
as opposed to Eq.~(\ref{h_CKG_h_nM_error}). This is a difference between the quasi-inertial (\ref{local_limit}) and the high acceleration limit (\ref{local_limit_alternative}) that is summarized in Table \ref{table_Rindler_limits}.

\section{Conclusions}\label{Conclusions}

We investigated the non-relativistic limit of scalar and Dirac particles in curved static spacetimes. It is well known that particles in flat spacetime are approximated by the same Schr\"{o}dinger equation in the non-relativistic limit [Eqs.~(\ref{Schrodinger_free_mass_positive_negative_frequencies}) and (\ref{Schrodinger_Dirac_free_mass_positive_negative_frequencies})]. On the contrary, scalar and Dirac fields in curved spacetimes have different non-relativistic asymptotic Hamiltonians $H_\text{S}$ and $H_\text{D}$. This implies that the two kinds of particles evolve differently when the gravitational field is sufficiently strong.

As an example, we considered non-relativistic particles in a Rindler metric with acceleration $\alpha$. For an $\alpha$ sufficiently large, $\Delta H=H_\text{D} - H_\text{S}$ cannot be ignored and lead to noticeable differences on the time evolution of the particles. If the spacetime is almost flat [Eq. (\ref{local_limit})], $\Delta H$ becomes negligible if compared to the gravitational potential $U_\text{g}$; in this way, one finds the usual Schr\"{o}dinger-Newton equation (\ref{Schrodinger_Rindler_scalar_local}) for both scalar and Dirac fields.

By considering higher orders, we find that GR corrections coming from $\Delta H $ are of order $ \epsilon^{3/2}$ [Eq.~(\ref{Delta_H_W_approximation_2})], while the GR corrections coming from the Klein-Gordon equation (\ref{Rindler_Klein_Gordon}) are of the order $\epsilon^2$ [Eq.~(\ref{H_S_h_CKG_error})]. This implies that an improved experimental precision will eventually unveil a second-order GR correction only for Dirac fields. We believe that this scaling addresses the possibility of observing spin-gravity coupling as a signal for general relativity in quantum particle phenomena.

We remark that the non-relativistic limit is often regarded as the one in which $c \rightarrow \infty$. However this limit may vary in a way dependent on the acceleration. Letting $a = \alpha / c^2$, the limit $c \rightarrow \infty$ does not specify if $\alpha$ has to go to infinity with finite $a$, or $a$ has to go to zero with finite $\alpha$. In the quasi-inertial limit (\ref{local_limit}), $\alpha$ is of order $\alpha \sim c^0$ and $a \sim c^{-2}$; as a result, the GR corrections are vanishing with respect to the Newtonian gravitational potential [Eq.~\ref{Delta_H_W_approximation_2}]. On the contrary, if $\alpha \sim c$ and $a \sim c^{-1}$ [Eq.~(\ref{local_limit_alternative})], the GR corrections are of the order of the Newtonian potential $ U_\text{g}$ [Eq.~(\ref{Delta_H_relativistic})], spoiling the difference between scalar and Dirac fields at the first non-trivial order.

\appendix

\section{}\label{appendix}

In this section, we give a proof of Eq.~(\ref{Klein_Gordon_curved_Dirac}) for any $\Psi$ that is solution of Eq.~(\ref{Dirac_curved}). Here we use the usual definition of derivatives $\nabla$ covariant with respect to the tensorial indexes $\mu, \nu, \rho, \sigma$ and $\mathcal{D}$ such that
\begin{equation}\label{fully_covatiant_derivative}
\mathcal{D}_\mu = \nabla_\mu + \Gamma_\mu.
\end{equation}
In this way, we replace Eq.~(\ref{Dirac_curved}) with
\begin{equation} \label{Dirac_curved_covariant}
\left( i c e_\alpha{}^\mu \gamma^\alpha \mathcal{D}_\mu  - \frac{m c^2}{\hbar} \right) \Psi = 0,
\end{equation}
Eq.~(\ref{spin_connection}) with
\begin{equation} \label{spin_connection_covariant}
\omega_{\alpha \beta \mu} = \eta_{\alpha \gamma} e^\gamma{}_\nu \nabla_\mu e_\beta{}^\nu,
\end{equation}
and Eq.~(\ref{Klein_Gordon_curved_Dirac}) with
\begin{equation} \label{Klein_Gordon_curved_Dirac_covariant}
\left[ c^2 g^{\mu \nu} \mathcal{D}_\mu \mathcal{D}_\nu - \left( \frac{mc^2}{\hbar}\right)^2 - \frac{c^2}{4} R \right] \Psi = 0.
\end{equation}
The aim here is to prove Eq.~(\ref{Klein_Gordon_curved_Dirac_covariant}) from Eq.~(\ref{Dirac_curved_covariant}).

We proceed by acting on the left of Eq.~(\ref{Dirac_curved_covariant}) with $i c e_\beta{}^\nu \gamma^\beta \mathcal{D}_\nu  + m c^2/\hbar$, in order to obtain
\begin{equation}\label{Dirac_curved_covariant_2}
\left[  -c^2  e_\beta{}^\nu \gamma^\beta \mathcal{D}_\nu \left( e_\alpha{}^\mu \gamma^\alpha \mathcal{D}_\mu \right)  -  \left(\frac{m c^2}{\hbar} \right)^2 \right] \Psi =  0.
\end{equation}
And thanks to Eqs.~(\ref{gamma_matrices_anticommutating_rule}), (\ref{spin_connection_covariant}) and the antisymmetry of $\omega_{\alpha \beta \mu} $ with respect to $\alpha$ and $\beta$ we prove that $\mathcal{D}_\nu$ and $e_\alpha{}^\mu \gamma^\alpha$ commute:
\begin{align}
& [\mathcal{D}_\nu, e_\alpha{}^\mu \gamma^\alpha] \nonumber \\ = & ( \nabla_\nu e_\alpha{}^\mu ) \gamma^\alpha + e_\alpha{}^\mu [\Gamma_\nu, \gamma^\alpha] \nonumber \\
= & \eta^{\beta \gamma} e_\gamma{}^\mu \omega_{\beta \alpha \nu} \gamma^\alpha - \frac{1}{4} e_\alpha{}^\mu \omega_{\beta \gamma \nu} [\gamma^\beta \gamma^\gamma , \gamma^\alpha]\nonumber \\
= &  e_\alpha{}^\mu \omega_{\beta \gamma \nu} \left( \eta^{\beta \alpha} \gamma^\gamma - \frac{1}{4}\gamma^\beta \gamma^\gamma \gamma^\alpha  + \frac{1}{4} \gamma^\alpha \gamma^\beta \gamma^\gamma \right)\nonumber \\
= &  e_\alpha{}^\mu \omega_{\beta \gamma \nu} \left( \frac{1}{2}\eta^{\beta \alpha} \gamma^\gamma - \frac{1}{4}\gamma^\beta \gamma^\gamma \gamma^\alpha  - \frac{1}{4} \gamma^\beta \gamma^\alpha \gamma^\gamma \right)\nonumber \\
= &  e_\alpha{}^\mu \omega_{\beta \gamma \nu} \left( \frac{1}{2}\eta^{\beta \alpha} \gamma^\gamma + \frac{1}{2}\eta^{\gamma \alpha} \gamma^\beta \right)\nonumber \\
= &  \frac{1}{2} e_\alpha{}^\mu ( \omega_{\beta \gamma \nu} + \omega_{ \gamma \beta \nu} ) \eta^{\beta \alpha} \gamma^\gamma \nonumber \\
= &  0 .
\end{align}
In this way Eq.~(\ref{Dirac_curved_covariant_2}) reads
\begin{equation}\label{Dirac_curved_covariant_3}
\left[ - c^2  e_\beta{}^\nu e_\alpha{}^\mu \gamma^\beta \gamma^\alpha \mathcal{D}_\nu \mathcal{D}_\mu  -  \left(\frac{m c^2}{\hbar} \right)^2 \right] \Psi =  0.
\end{equation}

We are now interested in the commutation relation $[ \mathcal{D}_\nu, \mathcal{D}_\mu ] \Psi$, which can be computed by separating $\mathcal{D}_\mu$ into $\nabla_\mu$ and $\Gamma_\mu$. Therefore, we derive the following quantities
\begin{subequations}\label{covariant_derivative_Gamma_commutating_rules}
\begin{align}
[ \nabla_\nu, \nabla_\mu ] \Psi = & \left( \partial_\nu \partial_\mu - \Gamma^\rho{}_{\nu \mu}\partial_\rho - \partial_\mu \partial_\nu + \Gamma^\rho{}_{\mu \nu}\partial_\rho \right) \Psi \nonumber \\
= & 0,
\end{align}
\begin{align}
[ \nabla_\nu, \Gamma_\mu ] \Psi = & ( \nabla_\nu \Gamma_\mu ) \Psi \nonumber \\
= & [ (\partial_\nu \Gamma_\mu) - \Gamma^\rho{}_{\nu \mu} \Gamma_\rho ]\Psi  \nonumber \\
= & \left[ -\frac{1}{4} (\partial_\nu \omega_{\alpha \beta \mu}) \gamma^\alpha \gamma^\beta - \Gamma^\rho{}_{\nu \mu} \Gamma_\rho \right]\Psi ,
\end{align}
\begin{align}
[ \Gamma_\nu, \Gamma_\mu ] = & \frac{1}{4} \omega_{\alpha \gamma \nu}  \omega_{\delta \beta \mu} [\sigma^{\alpha \gamma} , \sigma^{\delta \beta} ] \nonumber \\
 = & \frac{1}{4} \omega_{\alpha \gamma \nu}  \omega_{\delta \beta \mu} (-\eta^{\alpha \beta} \sigma^{\gamma \delta } + \eta^{\gamma \beta} \sigma^{\alpha \delta} \nonumber \\ 
 & + \eta^{\alpha \delta} \sigma^{\gamma \beta } - \eta^{\gamma \delta} \sigma^{\alpha \beta})\nonumber \\
 = & -\omega_{\alpha \gamma \nu}  \omega_{\delta \beta \mu} \eta^{\gamma \delta} \sigma^{\alpha \beta}\nonumber \\
 = & - \frac{1}{4}\omega_{\alpha \gamma \nu}  \omega_{\delta \beta \mu} \eta^{\gamma \delta} [\gamma^\alpha, \gamma^\beta ]\nonumber \\
 = & - \frac{1}{4} (\omega_{\alpha \gamma \nu}  \omega_{\delta \beta \mu} - \omega_{\alpha \gamma \mu}  \omega_{\delta \beta \nu} )\eta^{\gamma \delta} \gamma^\alpha \gamma^\beta,
\end{align}
\end{subequations}
where we have used the antisymmetry of spinorial indexes of $\omega_{\alpha \beta \mu}$ and the Clifford algebra commutation relation
\begin{equation}
[\sigma^{\alpha \gamma} , \sigma^{\delta \beta} ] = - \eta^{\alpha \beta} \sigma^{\gamma \delta } + \eta^{\gamma \beta} \sigma^{\alpha \delta} + \eta^{\alpha \delta} \sigma^{\gamma \beta } - \eta^{\gamma \delta} \sigma^{\alpha \beta}.
\end{equation}
From Eq.~(\ref{covariant_derivative_Gamma_commutating_rules}), one can derive
\begin{align}\label{covariant_derivative_Gamma_commutating_rules_2}
[ \mathcal{D}_\nu, \mathcal{D}_\mu ] \Psi  = & ( [ \nabla_\nu, \nabla_\mu ] + [ \nabla_\nu, \Gamma_\mu ] - [ \nabla_\mu, \Gamma_\nu ] + [ \Gamma_\nu, \Gamma_\mu ] ) \Psi\nonumber \\
 = &  -\frac{1}{4} [ (\partial_\nu \omega_{\alpha \beta \mu}) - (\partial_\mu \omega_{\alpha \beta \nu}) \nonumber \\
 & + (\omega_{\alpha \gamma \nu}  \omega_{\delta \beta \mu} - \omega_{\alpha \gamma \mu}  \omega_{\delta \beta \nu} )\eta^{\gamma \delta} ] \gamma^\alpha \gamma^\beta\Psi \nonumber \\
= &  -\frac{1}{4} e_\alpha{}^\rho e_\beta{}^\sigma R_{\rho \sigma \nu \mu} \gamma^\alpha \gamma^\beta\Psi, 
\end{align}
where
\begin{align}
R_{\rho \sigma \nu \mu} = & e^\alpha{}_\rho e^\beta{}_\sigma [ (\partial_\nu \omega_{\alpha \beta \mu}) - (\partial_\mu \omega_{\alpha \beta \nu}) \nonumber \\
 & + (\omega_{\alpha \gamma \nu}  \omega_{\delta \beta \mu} - \omega_{\alpha \gamma \mu}  \omega_{\delta \beta \nu} )\eta^{\gamma \delta}].
\end{align}
is the Riemann tensor in the form of Cartan's structure equation --- see for instance \cite{Gravitation}.

Such tensor has the following properties
\begin{equation}\label{Riemann_properties}
R_{\rho \sigma \nu \mu} = -R_{\sigma \rho \nu \mu} = - R_{\rho \sigma \mu \nu} = R_{\nu \mu \rho \sigma} = - R_{\rho \nu \mu \sigma} - R_{\rho \mu \sigma \nu}
\end{equation}
and is related to the Ricci scalar $R$ through the following identity
\begin{equation}\label{Riemann_properties_Ricci}
R = g^{\rho \nu} g^{\sigma \mu} R_{\rho \sigma \nu \mu}.
\end{equation}
Equations (\ref{Riemann_properties}) and (\ref{Riemann_properties_Ricci}) are used together with Eq.~(\ref{gamma_matrices_anticommutating_rule}) for the following chain of identities
\begin{align}
 & e_\beta{}^\nu e_\alpha{}^\mu  e_\gamma{}^\rho e_\delta{}^\sigma R_{\rho \sigma \nu \mu} \gamma^\beta \gamma^\alpha \gamma^\gamma \gamma^\delta \nonumber \\
= & -e_\beta{}^\nu e_\alpha{}^\mu  e_\gamma{}^\rho e_\delta{}^\sigma R_{\rho \sigma \nu \mu} \gamma^\beta \gamma^\alpha \gamma^\delta \gamma^\gamma \nonumber \\
= & e_\beta{}^\nu e_\alpha{}^\mu  e_\gamma{}^\rho e_\delta{}^\sigma R_{\rho \sigma \nu \mu} ( \gamma^\delta \gamma^\beta \gamma^\alpha + \gamma^\alpha \gamma^\delta \gamma^\beta ) \gamma^\gamma\nonumber \\
= & e_\beta{}^\nu e_\alpha{}^\mu  e_\gamma{}^\rho e_\delta{}^\sigma R_{\rho \sigma \nu \mu} ( -4\eta^{\delta \beta} \gamma^\alpha - \gamma^\beta \gamma^\delta \gamma^\alpha - \gamma^\alpha \gamma^\beta \gamma^\delta ) \gamma^\gamma\nonumber \\
= & e_\beta{}^\nu e_\alpha{}^\mu  e_\gamma{}^\rho e_\delta{}^\sigma R_{\rho \sigma \nu \mu} ( -4\eta^{\delta \beta} \gamma^\alpha + 2 \eta^{\delta \alpha} \gamma^\beta + \gamma^\beta \gamma^\alpha \gamma^\delta \nonumber \\
& - \gamma^\alpha \gamma^\beta \gamma^\delta ) \gamma^\gamma\nonumber \\
= & e_\beta{}^\nu e_\alpha{}^\mu  e_\gamma{}^\rho e_\delta{}^\sigma R_{\rho \sigma \nu \mu} ( 6 \eta^{\delta \alpha} \gamma^\beta + 2 \gamma^\beta \gamma^\alpha \gamma^\delta ) \gamma^\gamma\nonumber \\
= & e_\beta{}^\nu e_\alpha{}^\mu  e_\gamma{}^\rho e_\delta{}^\sigma R_{\rho \sigma \nu \mu} ( 6 \eta^{\delta \alpha} \gamma^\beta  \gamma^\gamma + 2 \gamma^\beta \gamma^\alpha \gamma^\delta  \gamma^\gamma)\nonumber \\
= & e_\beta{}^\nu e_\alpha{}^\mu  e_\gamma{}^\rho e_\delta{}^\sigma R_{\rho \sigma \nu \mu} [ 3  \eta^{\delta \alpha}( \gamma^\beta  \gamma^\gamma + \gamma^\gamma \gamma^\beta )\nonumber \\
& - 2 \gamma^\beta \gamma^\alpha  \gamma^\gamma\gamma^\delta ]\nonumber \\
= & e_\beta{}^\nu e_\alpha{}^\mu  e_\gamma{}^\rho e_\delta{}^\sigma R_{\rho \sigma \nu \mu} [ -6  \eta^{\delta \alpha} \eta^{\beta \gamma} - 2 \gamma^\beta \gamma^\alpha  \gamma^\gamma \gamma^\delta] \nonumber \\
= & -6 g^{\sigma \mu} g^{\rho \nu} R_{\rho \sigma \nu \mu} - 2 e_\beta{}^\nu e_\alpha{}^\mu  e_\gamma{}^\rho e_\delta{}^\sigma R_{\rho \sigma \nu \mu} \gamma^\beta \gamma^\alpha \gamma^\gamma \gamma^\delta \nonumber \\
= & -6 R - 2 e_\beta{}^\nu e_\alpha{}^\mu  e_\gamma{}^\rho e_\delta{}^\sigma R_{\rho \sigma \nu \mu} \gamma^\beta \gamma^\alpha \gamma^\gamma \gamma^\delta,
\end{align}
which leads to
\begin{equation}\label{Riemann_gamma_Ricci}
e_\beta{}^\nu e_\alpha{}^\mu  e_\gamma{}^\rho e_\delta{}^\sigma R_{\rho \sigma \nu \mu} \gamma^\beta \gamma^\alpha \gamma^\gamma \gamma^\delta = - 2 R.
\end{equation}

Equations (\ref{covariant_derivative_Gamma_commutating_rules_2}) and (\ref{Riemann_gamma_Ricci}) lead to the following identity
\begin{equation}
e_\beta{}^\nu e_\alpha{}^\mu \gamma^\beta \gamma^\alpha [ \mathcal{D}_\nu, \mathcal{D}_\mu ] \Psi = \frac{1}{2} R  \Psi,
\end{equation}
which in turn can be used in Eq.~(\ref{Dirac_curved_covariant_3}) together with Eq.~(\ref{gamma_matrices_anticommutating_rule}) in order to obtain Eq.~(\ref{Klein_Gordon_curved_Dirac_covariant}):
\begin{align}
&  - c^2  e_\beta{}^\nu e_\alpha{}^\mu \gamma^\beta \gamma^\alpha \mathcal{D}_\nu \mathcal{D}_\mu  \Psi \nonumber \\
 = &  - \frac{c^2}{2}  e_\beta{}^\nu e_\alpha{}^\mu \left( \{ \gamma^\beta, \gamma^\alpha \} + [ \gamma^\beta, \gamma^\alpha] \right) \mathcal{D}_\nu \mathcal{D}_\mu  \Psi \nonumber \\
= & \left( c^2  e_\beta{}^\nu e_\alpha{}^\mu \eta^{\beta \alpha } \mathcal{D}_\nu \mathcal{D}_\mu - \frac{c^2}{2}  e_\beta{}^\nu e_\alpha{}^\mu \gamma^\beta \gamma^\alpha [ \mathcal{D}_\nu , \mathcal{D}_\mu ]  \right) \Psi \nonumber \\
= & \left( c^2  g^{\nu \mu} \mathcal{D}_\nu \mathcal{D}_\mu - \frac{c^2}{4} R  \right) \Psi.
\end{align}

\section{} \label{Appendix_2}

A proof for Eq.~(\ref{F_tilde_Rindler_Hankel_2}) can be provided in the following way. Firstly, we manipulate Eq.~(\ref{F_tilde_Rindler}) by using the following identity \cite{FERREIRA2008223}
\begin{equation} \label{Hankel_Bessel}
K_\zeta (\xi) = i \frac{\pi}{2} \exp \left( i \frac{\pi}{2} \zeta \right) H_\zeta^{(1)} \left( e^{i \pi/2} \xi \right),
\end{equation}
where $H_\zeta^{(1)}$ is the Hankel function, $\zeta$ and $\xi$ are both complex values with $0 \leq \arg(\zeta) \leq \pi/2$ and $-\pi < \arg(\xi) \leq \pi$. Equation (\ref{Hankel_Bessel}) can be used in Eq.~(\ref{F_tilde_Rindler}) if we make the following identifications
\begin{align}\label{xi_zeta}
 & \zeta = e^{i\pi/2} \frac{\Omega}{c a}, & \xi = \sqrt{c^2 k_\perp^2 + \left(\frac{mc^2}{\hbar}\right)^2} \frac{e^{aX}}{c a}.
\end{align}
In this way, Eq.~(\ref{F_tilde_Rindler}) reads
\begin{align} \label{F_tilde_Rindler_Hankel}
& \tilde{F}(\Omega,\vec{k}_\perp,X) = \frac{e^{i \pi/2} }{4 \pi} \exp \left( -\frac{\pi}{2} e^{-i \pi/2} \zeta(\Omega) \right) \nonumber \\
& \times \sqrt{ \frac{\hbar}{a} \sinh \left( \pi e^{-i \pi/2} \zeta(\Omega) \right) }   H_{\zeta(\Omega)}^{(1)} \left( e^{i \pi/2} \xi(\vec{k}_\perp,X) \right),
\end{align}
where the functions $\zeta(\Omega)$ and $\xi(\vec{k}_\perp,X)$ are defined by Eq.~(\ref{xi_zeta}).

The limits (\ref{non_relativistic_limit_curved}) and (\ref{local_limit}) can be expressed in terms of $\zeta$ and $\xi$ in the following way
\begin{align}\label{local_limit_xi_zeta}
& e^{-i\pi/2} \zeta  \gg 1, & e^{i \pi/2} \xi \approx \zeta + \vartheta \zeta^{1/3},
\end{align}
with
\begin{equation}
\vartheta =    e^{i\pi/3} \left(\frac{m c}{\hbar a} \right)^{2/3} \left[ \frac{\hbar^2 k_\perp^2}{2 m^2 c^2} + a X - \left( \frac{\hbar \Omega}{m c^2} - 1  \right) \right],
\end{equation}
Equation (\ref{local_limit_xi_zeta}) can be proven in the following way:
\begin{align}
e^{-i\pi/2} \zeta = & \frac{\Omega}{c a}\nonumber \\
= & \frac{m c}{\hbar a} + \frac{m c}{\hbar a} \left( \frac{\hbar \Omega}{m c^2} - 1 \right)\nonumber \\
= & \frac{m c}{\hbar a} + \mathcal{O} \left( \epsilon^{-1/2} \right) \nonumber \\
= & \mathcal{O} \left( \epsilon^{-3/2} \right)
\end{align}
\begin{align}
\xi = & \frac{m c}{\hbar a} \left[ 1 + \frac{\hbar^2 k_\perp^2}{2 m^2 c^2} + \mathcal{O} \left( \epsilon^2  \right) \right]\left[ 1 + aX + \mathcal{O} \left( \epsilon^2  \right) \right] \nonumber \\
 = & \left[ \frac{\Omega}{ca} - \frac{m c}{\hbar a} \left( \frac{\hbar \Omega}{m c^2} - 1 \right) \right]  \nonumber \\
 & \times \left[ 1 + \frac{\hbar^2 k_\perp^2}{2 m^2 c^2} +  aX + \mathcal{O} \left( \epsilon^2  \right) \right]\nonumber \\
 = & \left\lbrace e^{-i\pi/2} \zeta - \left[ e^{-i\pi/2} \zeta +  \mathcal{O} \left( \epsilon^{-1/2}  \right) \right] \left( \frac{\hbar \Omega}{m c^2} - 1 \right) \right\rbrace  \nonumber \\
 & \times  \left[ 1 + \frac{\hbar^2 k_\perp^2}{2 m^2 c^2} +  aX + \mathcal{O} \left( \epsilon^2  \right) \right]\nonumber \\
 = &  e^{-i\pi/2} \zeta + e^{-i\pi/2} \zeta  \left[ \frac{\hbar^2 k_\perp^2}{2 m^2 c^2} +  aX  - \left( \frac{\hbar \Omega}{m c^2} - 1 \right) \right] \nonumber \\
 &  + \mathcal{O} \left( \epsilon^{1/2}  \right)\nonumber \\
 = &  e^{-i\pi/2} \zeta + \left( e^{-i\pi/2} \zeta \right)^{1/3} \left[ \frac{m c}{\hbar a} + \mathcal{O} \left( \epsilon^{-1/2}  \right) \right]^{2/3}  \nonumber \\
 & \times  \left[ \frac{\hbar^2 k_\perp^2}{2 m^2 c^2} +  aX  - \left( \frac{\hbar \Omega}{m c^2} - 1 \right) \right] + \mathcal{O} \left( \epsilon^{1/2}  \right)\nonumber \\
 = &  e^{-i\pi/2} \zeta + e^{-i\pi/6} \zeta^{1/3} \left( \frac{m c}{\hbar a} \right)^{2/3}  \nonumber \\
 & \times  \left[ \frac{\hbar^2 k_\perp^2}{2 m^2 c^2} +  aX  - \left( \frac{\hbar \Omega}{m c^2} - 1 \right) \right] + \mathcal{O} \left( \epsilon^0  \right)\nonumber \\
 = &  e^{-i\pi/2} \left( \zeta + \vartheta \zeta^{1/3} \right) + \mathcal{O} \left( \epsilon^0  \right).
\end{align}

When Eq.~(\ref{local_limit_xi_zeta}) holds, the limit of $H_\zeta^{(1)} \left( e^{i \pi/2} \xi \right)$ is \cite{FERREIRA2008223}
\begin{align} \label{limit_Hankel_Airy}
H_\zeta^{(1)} \left(  e^{i \pi/2} \xi \right) = & \frac{2^{4/3}}{\zeta^{1/3}} e^{-i\pi/3} \text{Ai} \left( -2^{1/3} e^{i2\pi/3} \vartheta \right) \nonumber \\
& + \mathcal{O} (|\zeta|^{-2/3}).
\end{align}
In terms of $\Omega$, $\vec{k}_\perp$ and $X$, Eq.~(\ref{limit_Hankel_Airy}) reads
\begin{align}
& H_{\zeta(\Omega)}^{(1)} \left(  e^{i \pi/2} \xi(\vec{k}_\perp, X) \right)\nonumber \\
 = & 2^{4/3} e^{-i \pi/2} \left( \frac{\Omega}{c a} \right)^{-1/3} \text{Ai} \left( 2^{1/3} \left(\frac{m c}{\hbar a} \right)^{2/3} \right. \nonumber \\
 & \left. \times \left[ \frac{\hbar^2 k_\perp^2}{2 m^2 c^2} + a X - \left( \frac{\hbar \Omega}{m c^2} - 1  \right) \right] \right) + \mathcal{O} \left( \epsilon  \right) \nonumber \\
 = & 2^{4/3} e^{-i \pi/2} \left\lbrace \frac{m c}{\hbar a} \left[ 1 + \mathcal{O} \left( \epsilon  \right) \right] \right\rbrace^{-1/3} \text{Ai} \left( 2^{1/3} \left(\frac{m c}{\hbar a} \right)^{2/3} \right. \nonumber \\
 & \left. \times\left[ \frac{\hbar^2 k_\perp^2}{2 m^2 c^2} + a X - \left( \frac{\hbar \Omega}{m c^2} - 1  \right) \right] \right)  + \mathcal{O} \left( \epsilon  \right)\nonumber \\
 = & 2^{4/3} e^{-i \pi/2} \left( \frac{\hbar a}{m c} \right)^{1/3} \text{Ai} \left( 2^{1/3} \left(\frac{m c}{\hbar a} \right)^{2/3} \right. \nonumber \\
 & \left. \times\left[ \frac{\hbar^2 k_\perp^2}{2 m^2 c^2} + a X - \left( \frac{\hbar \Omega}{m c^2} - 1  \right) \right] \right) \left[ 1 +  \mathcal{O} \left( \epsilon  \right) \right],
\end{align}
At the same time, in the limit $e^{-i \pi/2} \zeta \rightarrow \infty $,
\begin{equation}
\exp \left( -\frac{\pi}{2} e^{-i \pi/2} \zeta \right) \sqrt{ \sinh \left( \pi e^{-i \pi/2} \zeta \right) } \approx \frac{1}{\sqrt{2}}.
\end{equation}
Therefore, Eq.~(\ref{F_tilde_Rindler_Hankel}) can be approximated by Eq.~(\ref{F_tilde_Rindler_Hankel_2}).

\bibliography{bibliography}

\end{document}